\documentclass[prd,aps,nofootinbib,showpacs,showkeys,preprintnumbers]
{revtex4}
\usepackage{graphicx,epsf,amsmath,amsfonts,amssymb,amsbsy}
\usepackage{epsfig}
\newcommand{\ds}{\displaystyle}
\newcommand{\vev}[1]{\langle#1\rangle}
\newcommand{\mat}{\left ( \begin{array}}
\newcommand{\emat}{\end{array} \right )}
\newcommand{\vect}{\left ( \begin{array}{c}}
\newcommand{\evect}{\end{array} \right )}

\preprint{HU-EP-09/51}

\begin{document}

\title{ \bf Properties of the massive Gross-Neveu model with
nonzero baryon and isospin chemical potentials}
\author{D.~Ebert $^1$ and K. G.~Klimenko $^2$}
\affiliation{$^1$ Institute of Physics, Humboldt-University Berlin,
12489 Berlin, Germany\\
$^2$ IHEP and University of Dubna (Protvino branch), 142281
Protvino, Moscow Region, Russia}

\begin{abstract}
The properties of the two-flavored Gross-Neveu model with nonzero
current quark mass are investigated in the (1+1)-dimensional
spacetime at finite isospin $\mu_I$ as well as quark number $\mu$
chemical potentials and zero temperature. The consideration is
performed in the limit $N_c\to\infty$, i.e. in the case with an
infinite number of colored quarks. In the plane of parameters
$\mu_I,\mu$ a rather rich phase structure is found, which contains
phases with and without pion condensation. We have found a
great variety of one-quark excitations of these phases, including
gapless and gapped quasiparticles. Moreover, the mesonic mass
spectrum in each phase is also investigated.
\end{abstract}


\maketitle

\section{Introduction}

During the last decade great attention was paid to the investigation
of the QCD phase diagram in terms of baryonic as well as isotopic
(isospin) chemical potentials. First of all, this interest is
motivated by experiments on heavy-ion collisions, where we have to
deal with dense baryonic matter which has an evident isospin
asymmetry, i.e. different neutron and proton contents of initial
ions. Moreover, the dense hadronic/quark  matter inside compact
stars is also isotopically asymmetric. Generally speaking, one of
the important QCD applications is just to describe dense and hot
baryonic matter. However, in the above mentioned realistic
situations the density is rather small, and weak coupling QCD
analysis is not applicable. So, different nonperturbative methods or
effective theories such as chiral effective Lagrangians and
especially  Nambu -- Jona-Lasinio (NJL) type models \cite{njl} are
usually employed for the consideration of the properties of dense
and hot baryonic matter under heavy-ion experimental and/or compact
star conditions, i.e. in the presence of such external conditions as
temperature and chemical potentials, magnetic field, finite size
effects etc (see, e.g., the papers
\cite{2,asakawa,alford,klim,warringa,incera,ebert,vshivtsev}
and references therein). In particular, the color superconductivity
\cite{alford,klim} as well as parity vaiolation and charged pion
condensation \cite{son,frank,ek,jin,andersen,abuki} phenomena of
dense quark matter were investigated in the framework of these
QCD-like effective models.

It is necessary to note that an effective description of QCD in
terms of NJL models, i.e. through an employment of four-fermion
theories in (3+1)-dimensional space-time, is usually valid only at
{\it rather low} energies and densities. Besides, at present time
there is the consensus that another class of theories, the set of
(1+1)-dimensional Gross-Neveu (GN) type models \cite{gn,ft}, can
also be used for a reasonable qualitative consideration of the QCD
properties {\it without any restrictions} on the energy/density
values, which is in an encouraging contrast with NJL models. Indeed,
the GN type models are renormalizable, the asymptotic freedom and
spontaneous chiral symmetry breaking are another properties inherent
both for QCD and GN theories etc. In addition, the $\mu-T$ phase
diagram is qualitatively the same in QCD and GN model
\cite{wolff,kgk1,barducci,chodos,thies} (here $\mu$ is the quark
number chemical potential and $T$ is the temperature). Note also
that GN type models are suitable for the description of physics in
quasi one-dimensional condensed matter systems like polyacetylene
\cite{caldas}. Thus, due to the relative simplicity of GN models in
the leading order of the large $N_c$-expansion ($N_c$ is the number
of colored quarks), their usage is convenient for the application of
nonperturbative methods in quantum field theory \cite{okopinska}.

Before investigating different physical effects relevant to a real
(3+1)-dimensional world in the framework of two-dimensional GN
models, let us recall that there is a no-go theorem by
Mermin-Wagner-Coleman forbidding the spontaneous breaking of
continuous symmetries in two dimensions \cite{coleman}. This theorem
is based on the fact that in (1+1)-dimensional spacetime the Green
function (correlator) of two scalar fields has at large distances a
behavior $|x-y|^{-1/N_c}$. Thus, if we take the limit
$|x-y|\to\infty$ first, the correlator vanishes at finite $N_c$ and,
according to the cluster property, we formally obtain a zero vacuum
expectation value of the scalar field, i.e. a prohibition of
spontaneous symmetry breaking. However, there is a way to overcome
this no-go theorem. Indeed, if the limit $N_c\to\infty$ is taken
first, then for $|x-y|\to\infty$ we formally obtain a nonzero vacuum
expectation value for the scalar field, i.e. the possibility for
spontaneous symmetry breaking. It means that just the leading order
of the large $N_c$ approximation supplies us in any
(1+1)-dimensional model with a consistent field theory in which
spontaneous symmetry breaking might occur. At present time this fact
is well understood (see, e.g., the discussion in
\cite{barducci,chodos,thies}). This result restricts the range of
validity of the no-go theorem to the finite $N_c$-case only.
Clearly, since the no-go theorem does not work in the limit
$N_c\to\infty$, the investigation of any low-dimensional model in
the leading order at $N_c\to\infty$ is much more physically
appealing than the consideration of the model at finite $N_c$.

By this reason, such phenomena of dense QCD as color
superconductivity (spontaneous breaking of the color symmetry) or
charged pion condensation (spontaneous breaking of the continuous
isospin symmetry) might be simulated in terms of simpler
(1+1)-dimensional GN-type models in the leading order of the large
$N_c$ approximation (see, e.g., \cite{chodos} and \cite{ekzt},
correspondingly).

In our previous paper \cite{ekzt} the phase diagram of the
(1+1)-dimensional GN model with two massless quark flavors was
investigated under the constraint that quark matter occupies a
finite space volume (see also the relevant papers \cite{kim}). In
particular, the charged pion condensation phenomenon in cold quark
matter with zero baryonic density, i.e. at $\mu=0$, but nonzero
isotopic density, i.e. with nonzero isospin chemical potential
$\mu_I$, was studied there in the large $N_c$-limit. In contrast, in
the present paper we consider, in the leading order of the
$1/N_c$-expansion, the phase portrait of the above mentioned massive
GN model in a more general case, where, for simplicity, temperature
is taken to be zero, but both isospin and quark number chemical
potentials are nonzero, i.e. $\mu_I\ne 0$ and $\mu\ne 0$, and
spacetime is considered to have the usual topology, $R^1\times R^1$.
Our consideration is based on the case of homogeneous condensates
(an extension to inhomogeneous condensates in the case of $\mu_I=0$
was recently considered in \cite{thies,dunne}). We suppose that
these investigations will shed some new light on the physics of cold
dense and isotopically asymmetric quark matter which might exist in
compact stars, where baryon density is obviously nonzero (i.e.
$\mu\ne 0$).

The paper is organized as follows. In Sections II and III the
effective action and thermodynamic potential of the two-flavored
massive Gross-Neveu model are obtained in the presence of quark
number as well as isotopic chemical potentials. In Section IV the
phase structure of the model is investigated both in different
particular cases ($\mu\ne 0$, $\mu_I =0$ etc) and in the general case
of $\mu\ne 0$, $\mu_I\ne 0$. It turns out that at $\mu_I\ne 0$ and
rather small values of $\mu$, the gapped pion condensed phase (PC)
occurs. However, at larger values of $\mu$ several normal dense quark
matter phases (without PC) are found to exist with different
quasiparticle excitation properties of their ground states. In
Section V the meson mass spectrum of each phase is discussed. Some
technical details concerning the effective action and quark
propagator are relegated to two Appendices.

\section{ The model and its effective action}
\label{effaction}

We consider a (1+1)-dimensional model which describes dense quark
matter with two massive quark flavors ($u$ and $d$ quarks). Its
Lagrangian has the form
\begin{eqnarray}
&&  L=\bar q\Big [\gamma^\nu\mathrm{i}\partial_\nu-m_0
+\mu\gamma^0+\frac{\mu_I}2 \tau_3\gamma^0\Big ]q+ \frac {G}{N_c}\Big
[(\bar qq)^2+(\bar q\mathrm{i}\gamma^5\vec\tau q)^2 \Big ],
\label{1}
\end{eqnarray}
where the quark field $q(x)\equiv q_{i\alpha}(x)$ is a flavor
doublet ($i=1,2$ or $i=u,d$) and color $N_c$-plet
($\alpha=1,...,N_c$) as well as a two-component Dirac spinor (the
summation in (\ref{1}) over flavor, color, and spinor indices is
implied); $\tau_k$ ($k=1,2,3$) are Pauli matrices; the quark number
chemical potential $\mu$ in (\ref{1}) is responsible for the nonzero
baryonic density of quark matter, whereas the isospin chemical
potential $\mu_I$ is taken into account in order to study properties
of quark matter at nonzero isospin densities (in this case the
densities of $u$ and $d$ quarks are different).  Evidently, the
model (\ref{1}) is a simple generalization of the original
(1+1)-dimensional Gross-Neveu model \cite{gn} with a single massless
quark color $N_c$-plet to the case of two massive quark flavors and
additional chemical potentials. As a result, in the case under
consideration we have a modified flavor symmetry group, which depends
essentially on wether the bare quark mass $m_0$ and isospin chemical
potential $\mu_I$ take zero or nonzero values. Indeed, at $\mu_I
=0,m_0=0$ the Lagrangian (\ref{1}) is invariant under
transformations from the chiral $SU_L(2)\times SU_R(2)$ group.
Then, at $\mu_I \ne 0, m_0=0$ this symmetry is reduced to
$U_{I_3L}(1)\times U_{I_3R}(1)$, where $I_3=\tau_3/2$ is the third
component of the isospin operator (here and above the subscripts
$L,R$ mean that the corresponding group acts only on left, right
handed spinors, respectively). Evidently, this symmetry can also be
presented as $U_{I_3}(1)\times U_{AI_3}(1)$, where $U_{I_3}(1)$ is
the isospin subgroup and $U_{AI_3}(1)$ is the axial isospin
subgroup. Quarks are transformed under these subgroups as $q\to\exp
(\mathrm{i}\alpha\tau_3) q$ and $q\to\exp (\mathrm{i}
\alpha\gamma^5\tau_3) q$, respectively. In the case $m_0\ne
0,\mu_I=0$ the Lagrangian (\ref{1}) is invariant with respect to the
$SU_I(2)$, which is a diagonal subgroup of the chiral $SU_L(2)\times
SU_R(2)$ group. Finally, in the most general case with $m_0\ne
0,\mu_I\ne 0$ the initial model (\ref{1}) is symmetric under the
above mentioned isospin subgroup $U_{I_3}(1)$. In addition, in all
foregoing cases the model is color SU($N_c$) invariant.

The linearized version of the Lagrangian (\ref{1}), which contains
composite bosonic fields $\sigma (x)$ and $\pi_a (x)$ $(a=1,2,3)$,
has the following form:
\begin{eqnarray}
\tilde L\ds &=&\bar q\Big [\gamma^\nu\mathrm{i}\partial_\nu-m_0
+\mu\gamma^0+ \frac{\mu_I}2\tau_3\gamma^0-\sigma
-\mathrm{i}\gamma^5\pi_a\tau_a\Big ]q
 -\frac{N_c}{4G}\Big [\sigma\sigma+\pi_a\pi_a\Big ].
\label{2}
\end{eqnarray}
From the Lagrangian (\ref{2}) one gets the following constraint
equations for the bosonic fields
\begin{eqnarray}
\sigma(x)=-2\frac G{N_c}(\bar qq);~~~\pi_a (x)=-2\frac G{N_c}(\bar q
\mathrm{i}\gamma^5\tau_a q). \label{200}
\end{eqnarray}
Obviously, the Lagrangian (\ref{2}) is equivalent to the Lagrangian
(\ref{1}) when using the constraint equations (\ref{200}).
Furthermore, it is clear that the bosonic fields (\ref{200}) are
transforming under the isospin $U_{I_3}(1)$ subgroup in the
following manner:
\begin{eqnarray}
U_{I_3}(1):&&\sigma\to\sigma;~~\pi_3\to\pi_3;~~\pi_1\to\cos
(2\alpha)\pi_1+\sin (2\alpha)\pi_2;~~\pi_2\to\cos
(2\alpha)\pi_2-\sin (2\alpha)\pi_1, \label{201}
\end{eqnarray}
i.e the expression ($\pi_1^2+\pi_2^2$) remains unchanged under an
action of the isospin subgroup $U_{I_3}(1)$.

There is a common footing for obtaining both the thermodynamic
potential and one-particle irreducible Green functions of bosonic
$\sigma (x)$ and $\pi_a(x)$ fields (\ref{200}) which is based on the
effective action ${\cal S}_{\rm {eff}}(\sigma,\pi_a)$ of the model.
In the leading order of the large $N_c$-expansion (corresponding to
the one fermion-loop or mean field approximation), this quantity is
defined in terms of the Lagrangian (\ref{2}) through the relation
\begin{eqnarray}
\exp(\mathrm{i} {\cal S}_{\rm {eff}}(\sigma,\pi_a))=
  N'\int[d\bar q][dq]\exp\Bigl(\mathrm{i}\int\tilde  L\,d^2
  x\Bigr),\label{A1}
\end{eqnarray}
where $N'$ is a normalization constant. It is clear from (\ref{2})
and (\ref{A1}) that
\begin{eqnarray}
&&{\cal S}_{\rm {eff}} (\sigma,\pi_a)
=-N_c\int\frac{\sigma^2+\pi^2_a}{4G}d^2x+ \tilde {\cal S}_{\rm
{eff}}, \label{A2}
\end{eqnarray}
where the quark contribution to the effective action, i.e. the term
$\tilde {\cal S}_{\rm {eff}}$ in (\ref{A2}), is given by:
\begin{equation}
\exp(\mathrm{i}\tilde {\cal S}_{\rm {eff}})=N'\int [d\bar
q][dq]\exp\Bigl(\mathrm{i}\int\bar q D qd^2 x\Bigr)=\mbox{det}D.
\label{A3}
\end{equation}
Here we used the notations
\begin{eqnarray}
&&D=\mathrm{i}\gamma^\nu\partial_\nu- m_0+\mu\gamma^0+
\nu\tau_3\gamma^0-\sigma -\mathrm{i}\gamma^5\pi_a\tau_a \label{A4}
\end{eqnarray}
and $\nu=\mu_I/2$. Note also that $D$ is a nontrivial operator in
coordinate (x), spinor (s), and flavor (f) spaces, but it is
proportional to the unit operator in the $N_c$-dimensional color (c)
space. Then, using the general formula $\det D=\exp {\rm
Tr}_{xcfs}\ln D$, one obtains the following expression for the
effective action:
\begin{equation}
{\cal S}_{\rm {eff}}(\sigma,\pi_a)
=-N_c\int\frac{\sigma^2+\pi^2_a}{4G}d^2x -\mathrm{i}N_c{\rm
Tr}_{sfx}\ln D, \label{A5}
\end{equation}
where we have taken into account that the trace of the operator $\ln
D$ over the color space is proportional to $N_c$.

Starting from (\ref{A5}), one can define the thermodynamic potential
(TDP) of the model in the mean-field approximation:
\begin{equation}
{\cal S}_{\rm {eff}}~\bigg |_{~(\sigma,\pi_a=\rm {const})}
=-N_c\Omega_{\mu,\nu}(\sigma,\pi_a)\int d^2x. \label{A6}
\end{equation}
The ground state expectation values (mean values) of the bosonic
fields, $\vev{\sigma(x)}\equiv\sigma^o$ and $\vev{\pi_a(x)}
\equiv\pi_a^o$, are solutions of the gap equations for the TDP
$\Omega_{\mu,\nu} (\sigma,\pi_a)$ (in our approach all ground state
expectation values do not depend on coordinates $x$):
\begin{eqnarray}
\frac{\partial\Omega_{\mu,\nu}}{\partial\sigma}=0,~~~~~
\frac{\partial\Omega_{\mu,\nu}}{\partial\pi_a}=0,~~~~~
\mbox{where}~~~~a=1,2,3.\label{A7}
\end{eqnarray}
In particular, it follows from (\ref{A7}) that if $m_0\ne 0$ then
$\pi_3^o=0$. In addition, one can put $\pi_2^o=0$, since the
effective action depends on $\pi_1$ and $\pi_2$ fields through the
combination ($\pi_1^2+\pi_2^2$). Next, let us perform the following
shift of bosonic fields in (\ref{A5}):
$\sigma(x)\to\sigma(x)+\sigma^o$, $\pi_1(x)\to\pi_1(x)+\pi_1^o$,
whereas the other bosonic fields, $\pi_{2,3}$, stay unshifted.
(Obviously,  after shifting the new bosonic fields $\sigma (x),\pi_a
(x)$ now denote the small quantum fluctuations around the mean
values $\sigma^o,\pi^o_a$ of mesons rather than the original fields
(\ref{200})). Moreover, we use the notations $\sigma^o\equiv M-m_0$
and $\pi_1^o\equiv\Delta$. In this case
\begin{equation}
D=\Big (\mathrm{i}\gamma^\nu\partial_\nu- M+\mu\gamma^0+
\nu\tau_3\gamma^0 -\mathrm{i}\gamma^5\Delta\tau_1\Big )-\Big
(\sigma(x)+ \mathrm{i}\gamma^5\pi_a(x)\tau_a\Big )\equiv
S_0^{-1}-\Big(\sigma(x)+ \mathrm{i}\gamma^5\pi_a(x)\tau_a\Big ),
\label{A8}
\end{equation}
where $S_0$ is the quark propagator which is a 2$\times$2 matrix in
the flavor space, presented in Appendix \ref{ApE}. Then, expanding
the obtained expression into a Taylor-series up to second order of
small bosonic fluctuations $\sigma (x),\pi_a(x)$, we have
\begin{equation}
 {\cal S}_{\rm  {eff}}(\sigma,\pi_a)=
 {\cal S}_{\rm {eff}}^{(0)} +
 {\cal S}_{\rm  {eff}}^{(2)}(\sigma,\pi_a)
 +\cdots,
  \label{A9}
\end{equation}
where (due to the gap equations, the linear term in meson and
diquark fields is absent in (\ref{A9}))
\begin{eqnarray}
 \frac{1}{N_c}{\cal S}_{\rm {eff}}^{(0)}&&=-
 \int d^2x\frac{(M-m_0)^2+\Delta^2}{4G}
-\mathrm{i}{\rm Tr}_{sfx}\ln
 \left (S_0^{-1}\right )
\equiv-\Omega_{\mu,\nu}(M,\Delta)\int d^2x,\label{A10}\\
 \frac{1}{N_c}{\cal S}^{(2)}_{\rm
 {eff}}(\sigma,\pi_a)
 \!\!\!\!&&\!\!=
 -\int d^2x\frac{\sigma^2+\pi^2_a}{4G}
+\frac{\mathrm{i}}2{\rm Tr}_{sfx} \left\{S_0\Big(\sigma+
\mathrm{i}\gamma^5\pi_a\tau_a\Big )S_0\Big(\sigma+
\mathrm{i}\gamma^5\pi_a\tau_a\Big )\right\}.
  \label{A11}
\end{eqnarray}
The TDP $\Omega_{\mu,\nu}(M,\Delta)$ from (\ref{A10}) will be
calculated in the next section, where on the basis of this function
the phase structure of the GN model (1) in the leading order over
$1/N_c$ is considered. Note also that in (\ref{A9}) and (\ref{A11})
the bosonic fluctuation fields $\sigma,\pi_a$ are really the
coordinate dependent quantities. The trace of the $S_0$-operator
and the products of $\sigma,\pi_a$-fields in (\ref{A11}) should be
understood in the sense of formula (\ref{B1}) (see Appendix
\ref{ApB}). Note the remarkable property that the effective action
(\ref{A11}) is a generating functional of two-point and one-particle
irreducible (1PI) Green functions of $\sigma$- and $\pi$-mesons.
Indeed:
\begin{eqnarray}
&& \Gamma_{XY}(x-y)=-\frac{\delta^2{\cal S}^{(2)}_{\rm eff}}{\delta
Y(y)\delta X(x)},
  \label{A12}
\end{eqnarray}
where $X(x),Y(x)=\sigma (x),\pi_a(x)$ and $\Gamma_{XY}(x-y)$ is the
1PI Green function of the fields $X(x),Y(x)$. Variational
derivatives in (\ref{A12}) should be taken in accordance with the
general formula (\ref{B2}) (see Appendix \ref{ApB}). In the
following, on the basis of these Green functions we study the meson
mass spectrum in different phases of the model.

\section{ Thermodynamic potential}
\label{omega}

The Fourier transformation $\overline{S_0^{-1}}(p)$ of the inverse
quark propagator $S_0^{-1}$ (\ref{A8}) has the form:
\begin{eqnarray}
\overline{S_0^{-1}}(p)=\not\!p +\mu\gamma^0
+ \nu\tau_3\gamma^0-M-\mathrm{i}\gamma^5\Delta\tau_1.
\label{A13}
\end{eqnarray}
Clearly, in the direct product of spinor and flavor spaces it is a
4$\times$4 matrix, which has four eigenvalues:
\begin{eqnarray}
\epsilon_{1,2,3,4}=-M\pm\sqrt{(p_0+\mu)^2-p_1^2-\Delta^2+
\nu^2\pm 2\nu\sqrt{(p_0+\mu)^2-\Delta^2}}.
\label{A14}
\end{eqnarray}
Then, applying the general formula (\ref{B4}) to the expression
(\ref{A10}) for the thermodynamic potential, one gets:
\begin{eqnarray}
\Omega_{\mu,\nu}(M,\Delta)&=&\frac{(M-m_0)^2+\Delta^2}{4G}+
\mathrm{i}\sum_{i=1}^4\int\frac{d^2p}{(2\pi)^2}\ln (\epsilon_i
)\nonumber\\
&=&\frac{(M-m_0)^2+\Delta^2}{4G}+\mathrm{i}\int\frac{d^2p}{(2\pi)^2}
\ln\Big\{\Big [(p_0+\mu)^2-(E^+_{\Delta})^2\Big ]\Big
[(p_0+\mu)^2-(E^-_{\Delta})^2\Big ]\Big\}, \label{9}
\end{eqnarray}
where $E_\Delta^\pm=\sqrt{(E^\pm)^2+\Delta^2}$, $E^\pm=E\pm\nu$,
$\nu=\mu_I/2$ and $E=\sqrt{p_1^2+M^2}$. The system of
gap
equations directly follows from (\ref{9}):
\begin{eqnarray}
0=\frac{\partial\Omega_{\mu,\nu}(M,\Delta)}{\partial M}&\equiv&
\frac{M-m_0}{2G}-2\mathrm{i}M\int\frac{d^2p}{(2\pi)^2
E}\Big\{\frac{E^+}{(p_0+\mu)^2-(E^+_{\Delta})^2}+
\frac{E^-}{(p_0+\mu)^2-(E^-_{\Delta})^2} \Big\},\nonumber\\
0=\frac{\partial\Omega_{\mu,\nu} (M,\Delta)}{\partial\Delta}&\equiv&
\frac{\Delta}{2G}-2\mathrm{i}\Delta\int\frac{d^2p}{(2\pi)^2}\Big
\{\frac{1}{(p_0+\mu)^2-(E^+_{\Delta})^2}+\frac{1}{(p_0+\mu)^2-
(E^-_{\Delta})^2}
 \Big\}. \label{10}
\end{eqnarray}
The TDP $\Omega_{\mu,\nu}(M,\Delta)$ is symmetric under the
transformations $\mu\to -\mu$ and/or $\mu_I\to-\mu_I$. Hence, it is
sufficient to consider only the region $\mu\geq 0,\mu_I\geq 0$. In
this case, one can integrate in (\ref{9}) over $p_0$ with the help
of the formula
\begin{eqnarray*}
\int\frac{dp_0}{2\pi}\ln\Big [(p_0+a)^2-b^2\Big ]=
\frac{\mathrm{i}}2\Big\{|a-b|+|a+b|\Big\} \label{1000}
\end{eqnarray*}
(which is valid up to an infinite constant independent of
quantities $a$, $b$) and obtain:
\begin{eqnarray}
\Omega_{\mu,\nu}(M,\Delta)&&\!\!\!\!\!\!
=\frac{(M-m_0)^2+\Delta^2}{4G}-\int_{-\infty}^{\infty}\frac{dp_1}
{2\pi}\Big\{E^+_{\Delta}+E^-_{\Delta}+(\mu-E^+_{\Delta})\theta
(\mu-E^+_{\Delta})+(\mu-E^-_{\Delta})\theta
(\mu-E^-_{\Delta})\Big\}, \label{12}
\end{eqnarray}
where $\theta (x)$ is the Heaviside theta-function. In a similar
way, the system of gap equations (\ref{10}) is transformed to the
following one:
\begin{eqnarray}
0=\frac{\partial\Omega_{\mu,\nu}(M,\Delta)}{\partial M}&\equiv&
\frac{M-m_0}{2G}-M\int_{-\infty}^{\infty}\frac{dp_1}{2\pi
E}\Big\{\frac{\theta(E_\Delta^+-\mu)E^+}{E_\Delta^+}+
\frac{\theta(E_\Delta^--\mu)E^-}{E_\Delta^-} \Big\},\label{013}\\
0=\frac{\partial\Omega_{\mu,\nu} (M,\Delta)}{\partial\Delta}&\equiv&
\frac{\Delta}{2G}-\Delta\int_{-\infty}^{\infty}\frac{dp_1}{2\pi}\Big
\{\frac{\theta(E_\Delta^+-\mu)}{E_\Delta^+}+
\frac{\theta(E_\Delta^--\mu)}{E_\Delta^-} \Big\}. \label{13}
\end{eqnarray}
The coordinates (gap values) $M$ and $\Delta$ of the global minimum
point of the TDP (\ref{12}) supply us with two ground state
expectation values $\vev{\bar qq}$ and $\vev{\bar
q\mathrm{i}\gamma^5\tau_1 q}$, respectively, through the relations
$M=m_0+\vev{\sigma}$, $\Delta=\vev{\pi_1}$ and formulae (\ref{200}).
In particular, if the gap $\Delta$ is equal to zero, the ground
state of the model is isotopically symmetric and there is no
condensation of charged pions. However, if $\Delta\ne 0$, then the
ground state describes the phase with charged pion condensation,
where the isospin $U_{I_3}(1)$ symmetry is spontaneously broken. In
this phase the space parity is also spontaneously broken. Note also
that the physical essence of the other gap $M$ is the dynamical
quark mass which is not equal to the bare mass $m_0$, evidently.

It is clear that the TDP (\ref{12}) is an ultraviolet divergent
quantity, so one should renormalize it, using a special dependence
of the bare quantities such as the bare coupling constant $G$ and
the bare quark mass $m_0$ on the cutoff parameter $\Lambda$
($\Lambda$ restricts the integration region in the divergent
integrals, $|p_1|<\Lambda$). The renormalization procedure for the
simplest massive GN model was already discussed in the literature,
see, e.g., in \cite{kgk1,barducci,massive}. In a similar way, it is
easy to see that, cutting of the divergent integral in (\ref{12})
and using the substitution $G\equiv G(\Lambda)$ and $m_0\equiv
mG(\Lambda)$, where
\begin{eqnarray}
\frac{1}{2G(\Lambda)}=\frac{1}{\pi}\int_{-\Lambda}^\Lambda
dp_1\frac{1}{\sqrt{M_0^2+p_1^2}}=\frac{2}{\pi}\ln\left
(\frac{\Lambda+\sqrt{M_0^2+\Lambda^2}}{M_0}\right ) \label{16}
\end{eqnarray}
and $m$ is a new free finite renormalization--invariant massive
parameter \footnote{\label{foot1} Note, the quantity $m$ does not
equal to the physical or dynamical quark mass $M$. The last one is
defined by the pole position of the quark propagator. Alternatively,
it can be found as a gap, i.e. one of the coordinates of the global
minimum point of the thermodynamic potential (see also the remark in
the paragraph just after (\ref{13})).} (which does not depend on the
cutoff $\Lambda$), it is possible to obtain for the TDP (\ref{12}) a
finite renormalization--invariant expression. Namely,
\begin{eqnarray}
\Omega_{\mu,\nu}(M,\Delta)=\lim_{\Lambda\to\infty}\Omega_{\mu,\nu
}(M,\Delta;\Lambda), \label{17}
\end{eqnarray}
where
\begin{eqnarray}
\hspace{-0.3cm}\Omega_{\mu,\nu}(M,\Delta;\Lambda)
=\frac{M^2+\Delta^2}{4G(\Lambda)}-\frac{mM}{2}-\int_{-\Lambda}^
{\Lambda}
\frac{dp_1}{2\pi}\Big\{E^+_{\Delta}+E^-_{\Delta}+(\mu-E^+_{\Delta})
\theta(\mu-E^+_{\Delta})+(\mu-E^-_{\Delta})\theta
(\mu-E^-_{\Delta})\Big\}+\frac{\Lambda^2}{\pi}. \label{18}
\end{eqnarray}
(To obtain (\ref{18}) we have omitted the unessential constant
$\frac{m_0^2}{4G}$ as well as have added another one,
$\frac{\Lambda^2}{\pi}$.) In (\ref{16}) the cutoff independent
quantity $M_0$ is the dynamically generated quark mass in the
vacuum, i.e. at $\mu=0$ and $\mu_I=0$, taken in the chiral limit,
i.e. at $m_0=0$ (see below). (The renormalized expressions for the
gap equations are obtained in the limit $\Lambda\to\infty$, if the
replacements $G\to G(\Lambda)$, $m_0\to m G(\Lambda)$ and
$|p_1|<\Lambda$ are done in (\ref{13}), or by a direct
differentiation of the expression (\ref{17}).) The expression
(\ref{18}) can also be presented in the alternative form
\begin{eqnarray}
\Omega_{\mu,\nu}(M,\Delta;\Lambda)&&\!\!\!\!\!\!
=V_0(M,\Delta;\Lambda)-\frac{mM}{2}-\int_{-\Lambda}^{\Lambda}
\frac{dp_1}{2\pi}\Big\{E^+_{\Delta}+E^-_{\Delta}-2\sqrt{p_1^2+M^2+
\Delta^2}\nonumber\\
&&+(\mu-E^+_{\Delta})
\theta(\mu-E^+_{\Delta})+(\mu-E^-_{\Delta})\theta
(\mu-E^-_{\Delta})\Big\}, \label{19}
\end{eqnarray}
where
\begin{eqnarray}
V_0(M,\Delta;\Lambda)=\frac{M^2+\Delta^2}{4G(\Lambda)}-\frac{1}{\pi}
\int_{-\Lambda}^{\Lambda}
dp_1\sqrt{p_1^2+M^2+\Delta^2}+\frac{\Lambda^2}{\pi}. \label{20}
\end{eqnarray}
Obviously, the integral in (\ref{19}) is convergent at
$\Lambda\to\infty$. Since
\begin{eqnarray}
\lim_{\Lambda\to\infty}V_0(M,\Delta;\Lambda)=\frac{M^2+\Delta^2}{2\pi
}\left [\ln\left
(\frac{M^2+\Delta^2}{M_0^2}\right )-1\right ]\equiv V_0(M,\Delta),
\label{21}
\end{eqnarray}
one can easily obtain from (\ref{17}), (\ref{19}), and (\ref{21})
the following finite renormalization--invariant expression for the
TDP:
\begin{eqnarray}
\Omega_{\mu,\nu}(M,\Delta)&&\!\!\!\!\!\!
=V_0(M,\Delta)-\frac{mM}{2}-\int_{-\infty}^{\infty}
\frac{dp_1}{2\pi}\Big\{E^+_{\Delta}+E^-_{\Delta}-2\sqrt{p_1^2+M^2+
\Delta^2}\nonumber\\
&&+(\mu-E^+_{\Delta})
\theta(\mu-E^+_{\Delta})+(\mu-E^-_{\Delta})\theta
(\mu-E^-_{\Delta})\Big\}. \label{22}
\end{eqnarray}
Note that the integral in (\ref{22}) is convergent. In the
particular case of $\mu=0$, $\mu_I=0$ and $m=0$, i.e. for the
massless GN model in the vacuum, it follows from (\ref{22}):
\begin{eqnarray}
\Omega_{\mu,\nu}(M,\Delta){\Big |}_{\mu=0,\nu=0,m=0}
=\frac{M^2+\Delta^2}{2\pi}\left [\ln\left
(\frac{M^2+\Delta^2}{M_0^2}\right )-1\right ]. \label{23}
\end{eqnarray}
Since for a strongly interacting system the space--parity in the
vacuum is expected to be a conserved quantity, we put $\Delta$ equal
to zero in (\ref{23}). As a result, the global minimum of the TDP
(\ref{23}) lies in the point $M=M_0$, which means that in the vacuum
and at $m_0=0$ the dynamically generated quark mass is just the
parameter $M_0$ introduced in (\ref{16}). However, in the general
case, i.e. at nonzero values of the chemical potentials, the
dynamical quark mass depends certainly on $\mu,\mu_I$ and obeys the
system of the gap equations (\ref{013})-(\ref{13}) (or (\ref{10})).
Another free parameter of the massive GN model, the quantity $m$, is
not directly related to the quark mass, but rather to the mass of
$\pi$-mesons.

In the following, when studying the phase structure or the meson mass
spectrum, the quantity $M_0$ is still treated as a free parameter,
however the massive parameter of the model, $m\equiv\alpha M_0/\pi$,
is fixed by $\alpha=\alpha_0\approx 0.17$. In this case the vacuum
properties of the massive GN model resemble the situation in some
NJL-type models in realistic (3+1)-spacetime (for a more detailed
discussion, see the next section \ref{nu}).

For the forthcoming investigations we need also the expressions for
the density of quark number $n_q$ and isospin density $n_I$:
\begin{eqnarray}
n_q\equiv
-\frac{\partial\Omega_{\mu,\nu}}{\partial\mu}&&\!\!\!\!\!\!
=\int_{-\infty}^{\infty} \frac{dp_1}{2\pi}\Big\{
\theta(\mu-E^+_{\Delta})+\theta (\mu-E^-_{\Delta})\Big\},
\label{024}\\
n_I\equiv
-\frac{\partial\Omega_{\mu,\nu}}{2\partial\nu}&&\!\!\!\!\!\!= \frac
12\int_{-\infty}^{\infty} \frac{dp_1}{2\pi}\Big\{
\frac{E+\nu}{E^+_{\Delta}}\theta(E^+_{\Delta}-\mu)-\frac{E-\nu}
{E^-_{\Delta}}\theta (E^-_{\Delta}-\mu)\Big\}. \label{025}
\end{eqnarray}

\section{Phase structure of the model}
\subsection{Particular case: $\mu=0$, $\mu_I =0$}
\label{vacuum}

Introducing the notation $m\equiv\alpha M_0/\pi$, one can get from
(\ref{22}) the following expression for the TDP at $\mu=0$, $\mu_I
=0$ (usually, this quantity is called effective potential):
\begin{eqnarray}
\Omega_0(M,\Delta)=\frac{M^2+\Delta^2}{2\pi}\left [\ln\left
(\frac{M^2+\Delta^2}{M_0^2}\right )-1\right ]-\frac{\alpha
M_0M}{2\pi}. \label{230}
\end{eqnarray}
The corresponding gap equations look like:
\begin{eqnarray}
&&\frac{2\pi\partial\Omega_0(M,\Delta)}{\partial M}\equiv 2M\ln\left
(\frac{M^2+\Delta^2}{M_0^2}\right )-\alpha M_0=0, \label{231}\\
&&\frac{2\pi\partial\Omega_0(M,\Delta)}{\partial\Delta}\equiv
2\Delta\ln \left(\frac{M^2+\Delta^2}{M_0^2}\right )=0.\label{232}
\end{eqnarray}
The gap system (\ref{231})-(\ref{232}) has several solutions, but
the global minimum point (GMP) of the TDP (\ref{230}) corresponds to
the value $\Delta =0$. Then, at $\Delta =0$, the equation
(\ref{231}) vs $M$ has three solutions of different signs. Just the
one with largest absolute value corresponds to the GMP of the TDP.
This quantity (gap) is denoted by $M$ and depicted in Fig. 1 as a
function of the variable $\alpha$. Since the quark number density
$n_q$ and isospin density $n_I$ (\ref{024})-(\ref{025}) are equal to
zero in this GMP, we conclude that at $\mu=0$ and $\mu_I=0$ the
ground state of the model corresponds to the empty space, i.e. to
the vacuum. Hence, in this case the gap $M$ is the dynamical quark
mass in the vacuum. Clearly, the gap $M$ coincides with $M_0$ in the
chiral limit, $\alpha=0$. In addition, in Fig. 1 the behavior of the
$\pi$-meson mass $M_\pi$ vs $\alpha$ in the case of $\mu=0$ and
$\mu_I=0$ is also presented (it is the solution of the equation
(\ref{C11}) from section \ref{mesmas}). From the investigations of
section \ref{nu} it will become clear that $M_\pi$ coincides with
the critical value $\mu_{Ic}$ of the isotopical chemical potential
$\mu_I$, at which the system passes from the vacuum state to the
pion condensed phase. Just this fact is reflected in Fig. 1.
Moreover, we have also depicted in this figure the behavior of the
critical value $\mu_c$ vs $\alpha$ of the chemical potential $\mu$,
at which the system passes from the vacuum to the normal quark
matter phase at $\nu=0$ (see the next section \ref{mu}).

It is easily seen from Fig. 1 that the relation between the gap $M$
in the vacuum and the pion mass $M_\pi$ (at $\mu=0$ and $\mu_I=0$)
has a strong $\alpha$-dependency and for some values of this
parameter does not describe real physics. Recall, in real
(3+1)-dimensional physical models the dynamical quark mass $M$ is
usually greater than $M_\pi$ at $\mu=0$ and $\mu_I=0$ and depends on
the model parameters (coupling constants, cutoff parameter etc). In
particular, the values $M=350$ MeV and $M_{\pi}=140$ MeV, i.e.
$M/M_{\pi}=5/2$, are often used in the NJL-model investigations of
dense quark matter \cite{eky}. So, in the following consideration of
the phase structure of the model (1) and its meson mass spectrum in
the most general case of $\mu\ne 0$ and $\nu\ne 0$, we will suppose
the same relation between $M$ and $M_\pi$ at $\mu=0$ and $\mu_I=0$.
Evidently (see Fig. 1), this choice corresponds to
$\alpha=\alpha_0\approx 0.17$. Having fixed the parameter
$\alpha=\alpha_0\approx 0.17$, it is then possible to obtain
$M/M_0\approx 1.04$, $M_\pi/M_0\approx 0.42$ and $m/M_0\approx
0.05$, where $M_0$ is the dynamical quark mass in the massless GN
model at $\mu=0$ and $\mu_I=0$.
\begin{figure}
 \includegraphics[width=0.45\textwidth]{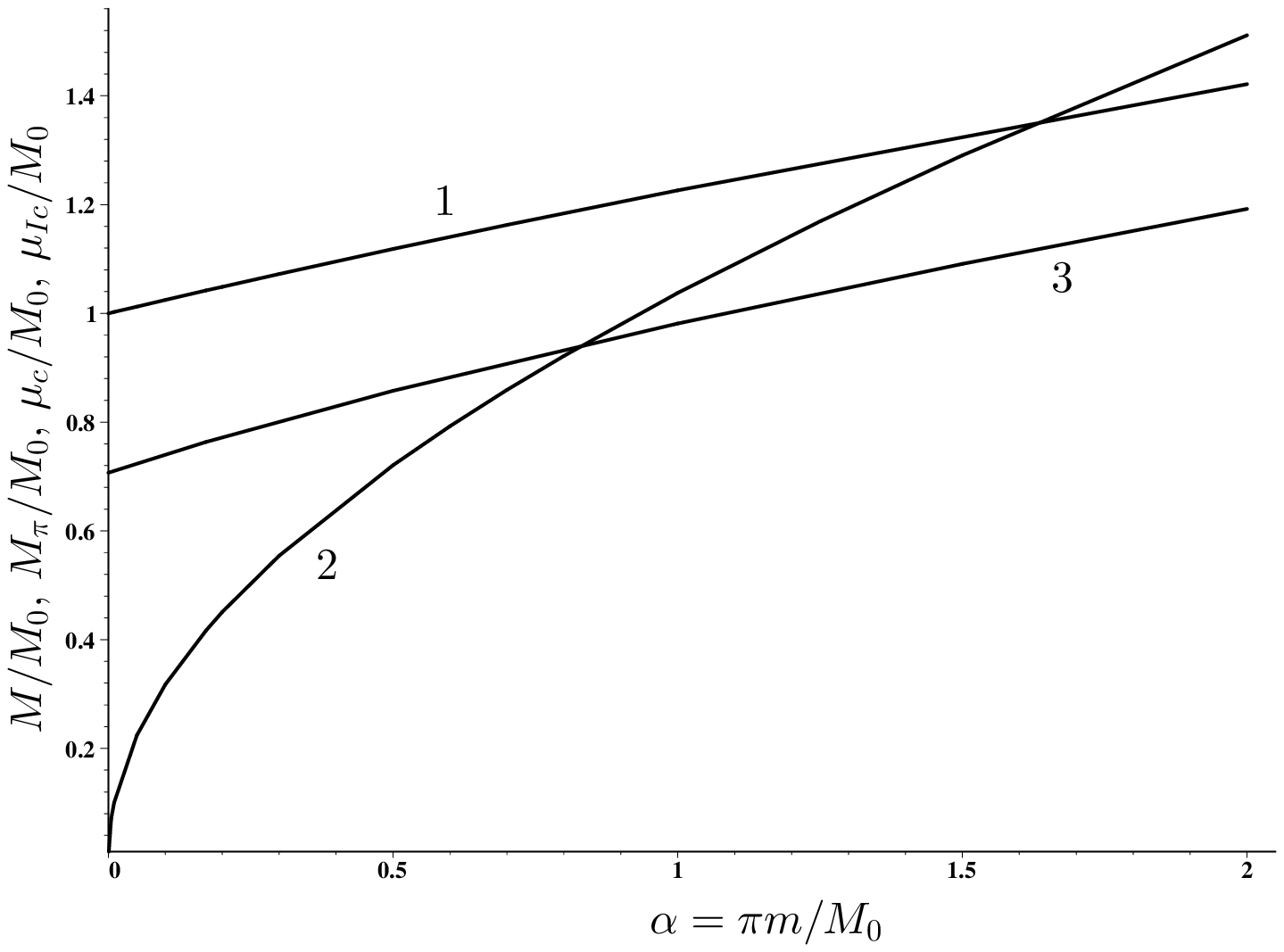}
 \hfill
 \includegraphics[width=0.45\textwidth]{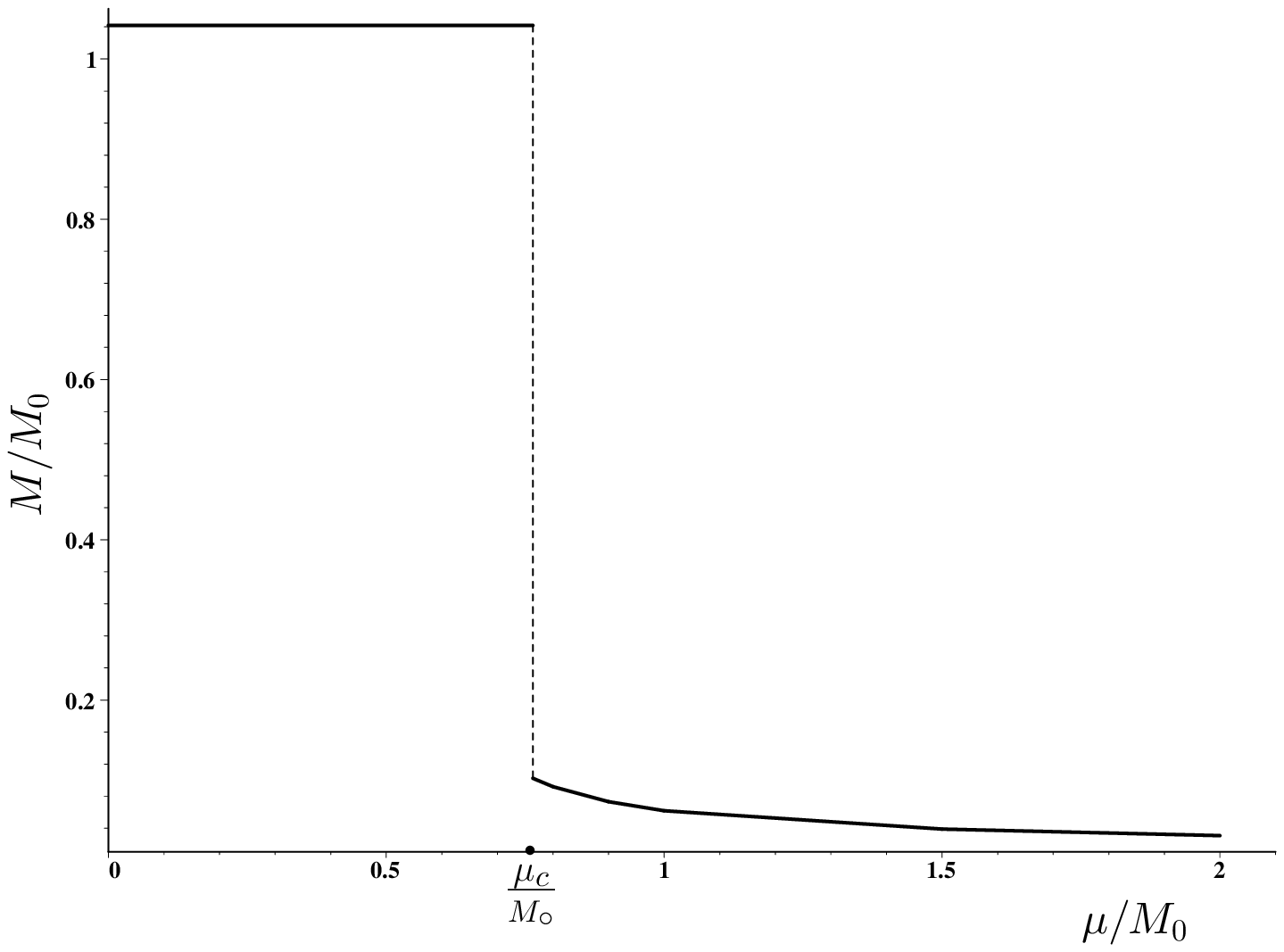}\\
\parbox[t]{0.45\textwidth}{
 \caption{Dynamical quark mass $M$ (curve 1)
 and $\pi$-meson mass $M_\pi$ (curve 2) vs
 $\alpha\equiv\pi m/M_0$ at $\mu=0$, $\mu_I=0$. The curve 3 is the
 critical value $\mu_c$ of the vacuum -- normal quark matter phase
 transition (at $\mu_I=0$); the critical value $\mu_{Ic}$ of the
 vacuum -- PC phase transition is also given by
the curve 2, i.e. $\mu_{Ic}=M_\pi$ (see in the section \ref{nu}).}
 }\hfill
\parbox[t]{0.45\textwidth}{
\caption{Dynamical quark mass $M$ vs
 $\mu$ at $\mu_I=0$ and
$\alpha=\alpha_0\approx 0.17$. Here $\mu_c/M_0\approx 0.76$.}  }
\end{figure}

\subsection{Particular case: $\mu\ne 0$, $\mu_I= 0$}
\label{mu}

Using again the notation $m\equiv\alpha M_0/\pi$, one can get from
(\ref{22}) the following expression for the TDP at $\mu\ne 0$,
$\mu_I =0$:
\begin{eqnarray}
&&\Omega_\mu(M,\Delta)=\frac{M^2+\Delta^2}{2\pi}\left [\ln\left
(\frac{M^2+\Delta^2}{M_0^2}\right )-1\right ]-\frac{\alpha
M_0M}{2\pi}\nonumber\\
&&+\frac{\theta(\mu-\sqrt{M^2+\Delta^2})}{\pi} \left
[(M^2+\Delta^2)\ln\left
(\frac{\mu+\sqrt{\mu^2-M^2-\Delta^2}}{\sqrt{M^2+\Delta^2}} \right
)-\mu\sqrt{\mu^2-M^2-\Delta^2}\right ]. \label{25}
\end{eqnarray}
It follows from the gap equations for the TDP (\ref{25}) that
$\Delta =0$ in its global minimum point, whereas the $M$-coordinate
of the GMP obeys the equation:
\begin{eqnarray}
&&\theta (\mu^2-M^2)\ln\frac{(\mu+\sqrt{\mu^2-M^2})^2}{M^2_0}+
\theta(M^2-\mu^2)\ln\frac{M^2}{M_0^2}=\frac{\alpha
M_0}{2M}.\label{26}
\end{eqnarray}
Studying the GMP of the TDP (\ref{25}) with the help of the
stationary equation (\ref{26}), it is possible to show that at
$\mu<\mu_c$ the GMP is arranged in the point $(M,\Delta=0)$, where
both the critical value $\mu_c$ and the gap $M$ are depicted in Fig.
1. In this case the system is arranged in the vacuum state with
$n_q=0$ and $n_I=0$. However, if $\mu>\mu_c$ then the phase which is
usually called the normal quark matter phase is realized in the
model. In this phase the quark number density $n_q$ is nonzero,
however the isospin density $n_I=0$ at $\mu_I=0$. In the particular
case with $\alpha =\alpha_0\approx 0.17$ the behavior of the
$M$-coordinate (gap) of the GMP is presented in Fig. 2, where
$\mu_c\approx 0.76 M_0$, as a function of $\mu$.
\begin{figure}
\includegraphics[width=0.45\textwidth]{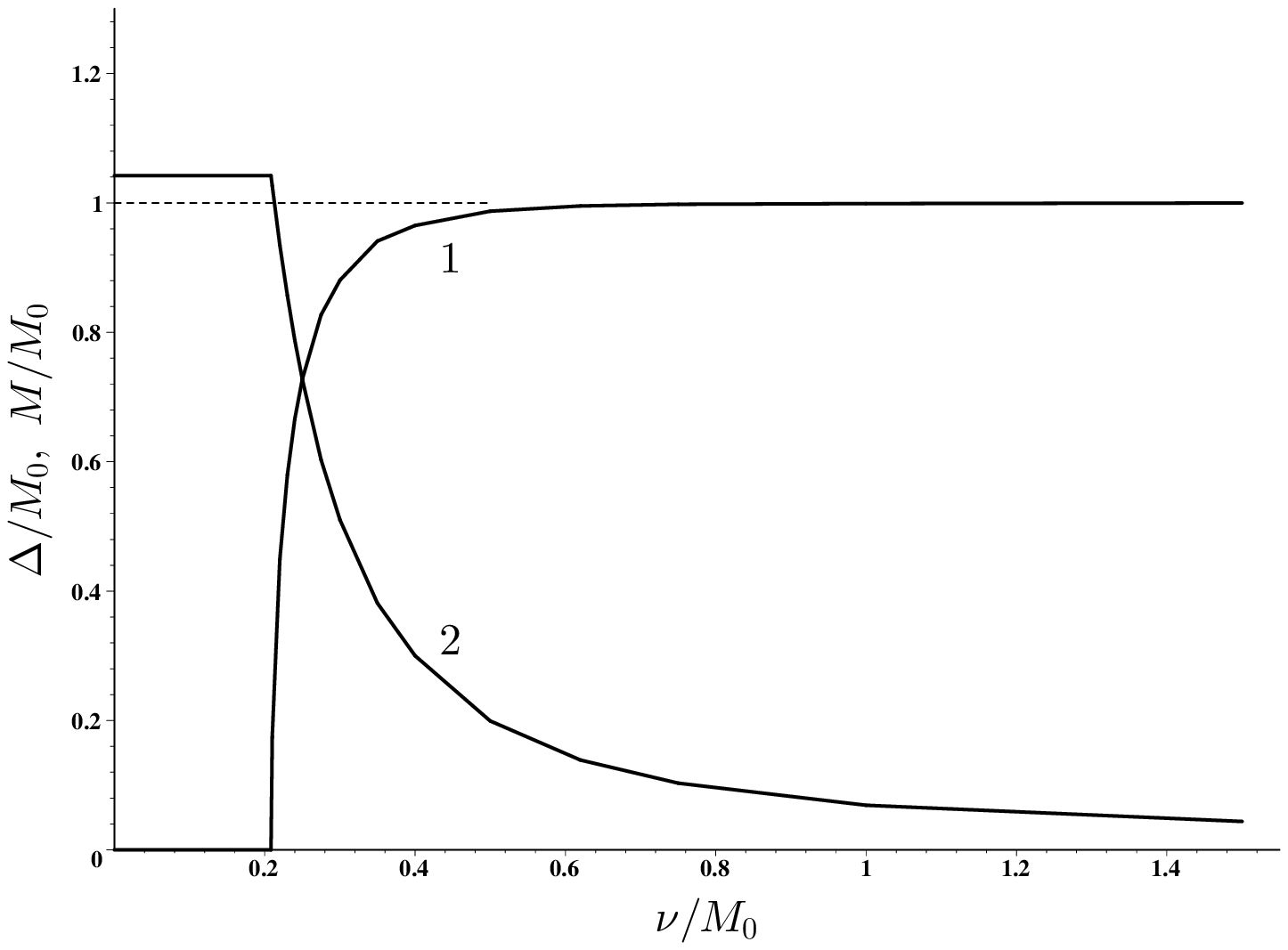}
\hfill
\includegraphics[width=0.45\textwidth]{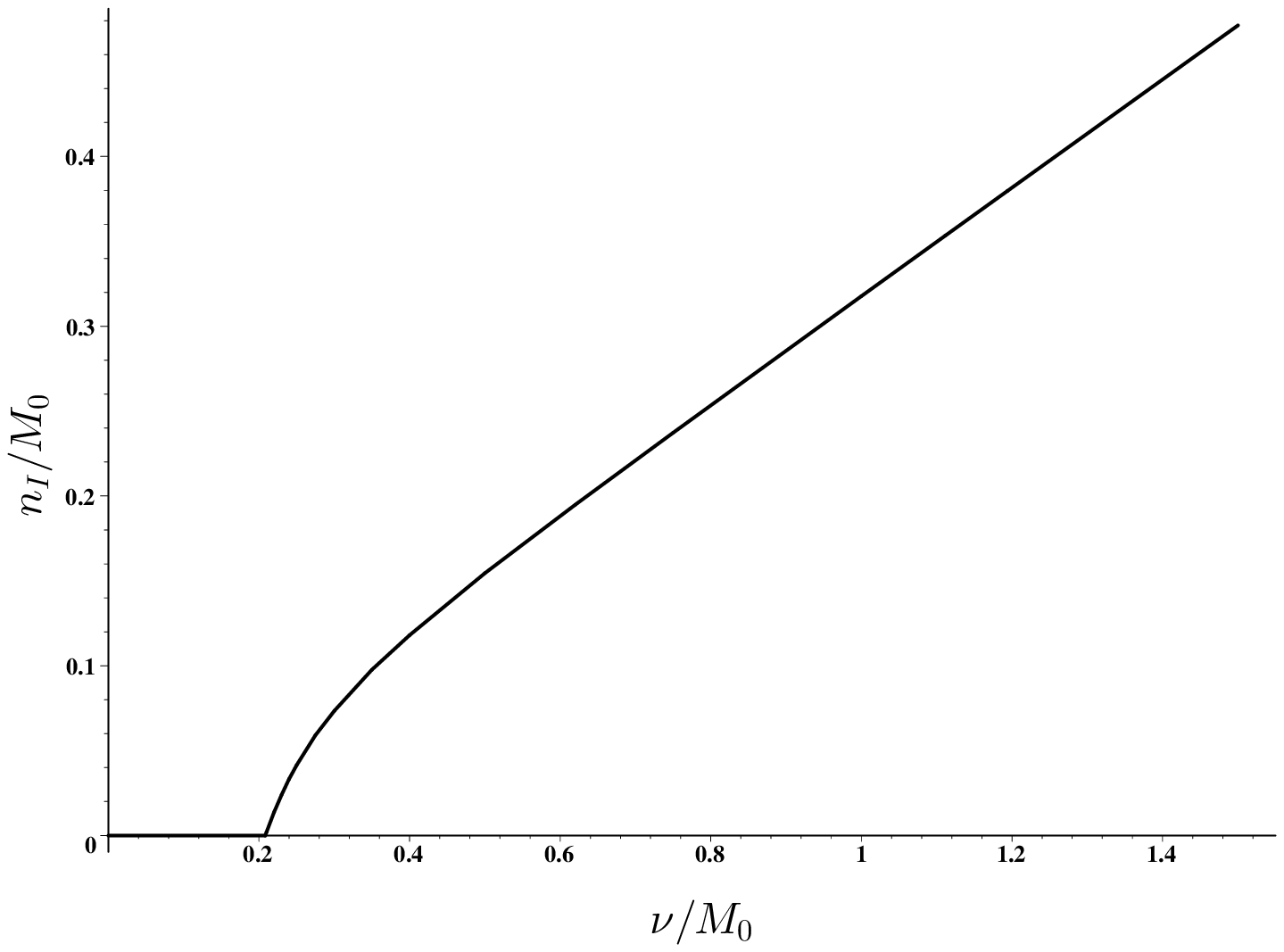}\\
\parbox[t]{0.45\textwidth}{
 \caption{The gaps $\Delta$ (curve 1) and $M$
(curve 2) vs $\nu\equiv\mu_I/2$ in the case $\mu=0$ and
$\alpha=\alpha_0\approx 0.17$.}  } \hfill
\parbox[t]{0.45\textwidth}{
\caption{Isospin density $n_I$ vs $\nu$ at
 $\mu=0$ and $\alpha=\alpha_0\approx 0.17$ in the vacuum (at
 $\nu<\nu_c\approx 0.21M_0$), where $n_I\equiv 0$, and in the PC
 phase (at $\nu>\nu_c$), where $n_I\ne 0$.}  }
\end{figure}

\subsection{Particular case: $\mu=0$, $\mu_I\ne 0$}
\label{nu}

In this case the TDP (\ref{22}) has the following form:
\begin{eqnarray}
\Omega_{\nu}(M,\Delta)&&\!\!\!\!\!\!
=V_0(M,\Delta)-\frac{\alpha M_0M}{2\pi}-\int_{-\infty}^{\infty}
\frac{dp_1}{2\pi}\Big\{E^+_{\Delta}+E^-_{\Delta}-2\sqrt{p_1^2+M^2+
\Delta^2}\Big\}. \label{24}
\end{eqnarray}
The corresponding system of the gap equations looks like:
\begin{eqnarray}
&&\frac{2\pi\partial\Omega_{\nu}(M,\Delta)}{\partial M}\equiv
2M\ln\left (\frac{M^2+\Delta^2}{M_0^2}\right )-\alpha
M_0-2M\int_{0}^{\infty}
dp_1\Big\{\frac{E^+}{EE^+_{\Delta}}+\frac{E^-}{EE^-_{\Delta}}-
\frac{2}{\sqrt{p_1^2+M^2+\Delta^2}}\Big\}=0, \label{240}\\
&&\frac{2\pi\partial\Omega_{\nu}(M,\Delta)}{\partial\Delta}\equiv
2\Delta \left [\ln\left (\frac{M^2+\Delta^2}{M_0^2}\right
)-\int_{0}^{\infty}
dp_1\Big\{\frac{1}{E^+_{\Delta}}+\frac{1}{E^-_{\Delta}}-
\frac{2}{\sqrt{p_1^2+M^2+\Delta^2}}\Big\}\right ]=0, \label{241}
\end{eqnarray}
where we have used the notations adopted after formula (\ref{9}).
The coordinates (gap values) $M\equiv M(\nu)$ and
$\Delta\equiv\Delta(\nu)$ of the global minimum point of the TDP
(\ref{24}) obey the gap equations (\ref{240}) and (\ref{241}). (In
the present section we find it convenient to stress explicitly the
fact that the GMP is indeed a function of the parameter $\nu$.)

Recall the situation in (3+1)-dimensional NJL models with pion
condensation, if the bare (current) quark mass is nonzero
\cite{jin}. In this case at some critical value $\mu_{Ic}$ of the
isospin chemical potential, which is just the pion meson mass
$M_{\pi}$ in the vacuum at $\mu=0$ and $\mu_I=0$, i.e.
$\mu_{Ic}=M_{\pi}$, there is a continuous 2nd order phase transition
from the vacuum phase (which is realized at $\nu <\nu_c=M_\pi/2$)
with $M(\nu)\equiv M(0)\ne 0$, $\Delta (\nu)=0$ to the pion
condensed one (at $\nu >\nu_c$), where $M(\nu)\ne 0$, $\Delta
(\nu)\ne 0$. This means that the TDP global minimum point
$(M(\nu),\Delta (\nu))$, corresponding to the pion condensed phase,
has the following property: $M(\nu)\to M(\nu_c)\equiv M(0)$,
$\Delta(\nu)\to 0$, if $\nu\to\nu_{c+}$. Here we again use the
notations $\nu=\mu_{I}/2$ as well as $M(0)$ for the dynamical quark
mass in the vacuum.

It turns out that the same  qualitative picture of the pion
condensed phase transition occurs in the framework of the massive GN
model. Indeed, numerical investigations of the TDP (\ref{24}) show
that at some critical point $\nu_c$ there is a second order phase
transition from the vacuum phase to the phase with charged pion
condensation. It means that the GMP of the TDP (\ref{24}) is a
continuous function vs $\nu$ in the critical point $\nu=\nu_c$. Now,
in order to define $\nu_c$ and to prove that the equality
$\nu_c=M_\pi/2$ is also valid in the case of the massive GN model,
it is necessary to remark that at $\nu>\nu_c$ the coordinates
$(M(\nu),\Delta (\nu))$ of the GMP of the TDP (\ref{24}) convert the
expression in the square brackets of (\ref{241}) into zero.
Moreover, the equation (\ref{240}) is also fulfilled. Since at
$\nu=\nu_c$ we have a continuous phase transition, i.e. $\Delta
(\nu_c)=0$, $M(\nu_c)\equiv M(0)$,
\footnote{The quantity $M(0)$ vs $\alpha$ is nothing else than the
gap $M$ depicted in Fig. 1 as curve 1.}
in the critical point $\nu=\nu_c$ this pair of equations is
transformed into the following one
\begin{eqnarray}
\alpha M_0&=&2M(0)\ln\frac{M^2(0)}{M_0^2}, \label{242}\\
\ln\frac{M^2(0)}{M_0^2}&=&2\nu^2_c\int_{0}^{\infty}
dp_1\frac{1}{\sqrt{p_1^2+M^2(0)}(p_1^2+M^2(0)-\nu^2_c)}.
\label{243}
\end{eqnarray}
Next, by inserting equation (\ref{243}) into the right hand side
of (\ref{242}), we find the useful relation
\begin{eqnarray}
\frac{\alpha M_0}{2M(0)}=2\nu^2_c\int_{0}^{\infty}
dp_1\frac{1}{\sqrt{p_1^2+M^2(0)}(p_1^2+M^2(0)-\nu^2_c)}.
\label{244}
\end{eqnarray}
In the next sections we will study the meson masses in different
phases of the model. In particular, we shall there derive an
equation for the $\pi$-meson mass $M_\pi$ in the vacuum at $\mu=0$
and $\mu_I=0$ (comp.(\ref{C11})). Comparing (\ref{244}) with this
equation, it follows that $\nu_c=M_\pi/2$, i.e. the critical value
$\mu_{Ic}$ is equal to the $\pi$-meson mass $M_\pi$ at $\mu=0$ and
$\mu_I=0$ for arbitrary values of $\alpha$. (Of course, one should
take into account that the corresponding dynamical quark mass $M$
appearing in this equation is nothing else than the parameter $M(0)$
of the present section.) As a result, the dependence of $\mu_{Ic}$
and $M_\pi$ vs $\alpha$ is presented by the same curve 2 of Fig. 1.

Clearly, at $\nu<\nu_c$ we have a phase which corresponds to the
empty space (here both $n_q$ and $n_I$ are equal to zero). Due to
this property, we use the notation vacuum for this phase.
\footnote{\label{vacrem} By definition, the vacuum is here the phase
with zero densities $n_q$ and $n_I$. However, one should realize
that in a most general case the (dynamical) properties of its ground
state depend on the values of $\mu$ and $\mu_I$. Indeed, in the
model under consideration at $\mu_I=0$ there is an $SU_I(2)$
symmetry of the ground state in the vacuum phase. As a result, all
three pions have a common mass. However, at $\mu_I\ne 0$, i.e. when
the ground state symmetry is reduced to the $U_{I_3}(1)$ subgroup,
$\pi$-mesons have different masses in this phase (comp. section
\ref{mass1}).} In the vacuum phase one has $\Delta=0$, but the gap
$M$ is nonzero and does not depend on $\nu$ (its behavior vs
$\alpha$ is shown in Fig. 1). At $\nu>\nu_c$ the pion condensation
(PC) phase with $n_q=0$ and $n_I\ne 0$ is realized in the model.
Inside this phase both gaps $M$ and $\Delta$ are nonzero and depend
on $\nu$. The isospin $U_{I_3}(1)$ symmetry is spontaneously broken
in the PC phase. For the particular parameter value
$\alpha=\alpha_0\approx 0.17$ the behavior of gaps vs $\nu$ is shown
in Fig. 3, where $\nu_c\approx 0.21 M_0$. In Fig. 4 the isospin
density $n_I$ vs $\nu$ is presented.

\subsection{General case: $\mu\ne 0$, $\mu_I\ne 0$}
\label{general}

In this case, starting from the TDP (\ref{22}) we obtain the
following gap equations:
\begin{eqnarray}
\frac{2\pi\partial\Omega_{\mu,\nu}(M,\Delta)}{\partial
M}&\equiv&2M\ln\left
(\frac{M^2+\Delta^2}{M_0^2}\right )-\alpha M_0\nonumber\\
&&-2M\int_{0}^{\infty}
dp_1\Big\{\frac{E^+}{EE^+_{\Delta}}+\frac{E^-}{EE^-_{\Delta}}-
\frac{2}{\sqrt{E^2+\Delta^2}}-\frac{E^+\theta(\mu-E^+_{\Delta})
}{EE^+_{\Delta}}-\frac{E^-\theta(\mu-
E^-_{\Delta})}{EE^-_{\Delta}}\Big\}=0, \label{27}\\
\frac{2\pi\partial\Omega_{\mu,\nu}(M,\Delta)}{\partial\Delta}&\equiv&
2\Delta\Big [\ln\left (\frac{M^2+\Delta^2}{M_0^2}\right
)\nonumber\\&&-\int_{0}^{\infty}
dp_1\Big\{\frac{1}{E^+_{\Delta}}+\frac{1}{E^-_{\Delta}}-
\frac{2}{\sqrt{E^2+\Delta^2}}-
\frac{\theta(\mu-E^+_{\Delta})}{E^+_{\Delta}}-\frac{\theta(\mu-
E^-_{\Delta})}{E^-_{\Delta}}\Big\} \Big ]=0. \label{28}
\end{eqnarray}
Based on these equations, we have studied the properties of the GMP
of the TDP (\ref{22}) in the particular case of
$\alpha=\alpha_0\approx 0.17$ and found the phase portrait,
presented in Fig. 5. There the vacuum, pion condensation as well as
three normal quark matter phases I, II and III are arranged. In the
pion condensation phase the gaps $\Delta$ and $M$ are nonzero, so
here the isospin $U_{I_3}(1)$ symmetry is broken spontaneously.
Throughout of this phase the gaps $\Delta$ and $M$ do not depend on
$\mu$. It turns out that their dependencies on $\nu$ in the PC phase
at $\mu\ne 0$ are the same as in the PC phase at $\mu=0$ (see Fig.
3). For points $(\nu,\mu)$, taken from the other phases of Fig. 5,
the $\Delta$-coordinate of the global minimum point of the TDP is
zero (as a result, in these phases the isospin $U_{I_3}(1)$ symmetry
remains intact), but the $M$-coordinate of GMP is not zero. Namely,
inside the vacuum phase the gap $M$ does not depend on $(\nu,\mu)$,
i.e. it is a constant. (In particular, here $M\approx 1.04M_0$ at
$\alpha=\alpha_0 \approx 0.17$.) Our analysis shows that on the
boundary between the vacuum and pion condensation phases the gaps
are continuous functions vs $\mu$ and $\nu$. Hence, we conclude that
a transition from the vacuum to the pion condensed phase or
conversely is a second order one.

It turns out that the gap $M$ is a continuous $(\mu,\nu)$-function
inside each of domains I, II, and III of Fig. 5. In contrast, it is
changed by a jump when each boundary between I, II, and III phases
is crossed. To become convinced in this, look at Fig. 6, where the
behavior of $M$ vs $\nu$ at two different fixed values of $\mu$ is
presented (there, the phase II is shrank to the interval $(a_1,b_1)$
at $\mu=0.84M_0$ and to the interval $(a_2,b_2)$ at $\mu=0.94M_0$).
As a result, we conclude that on these boundaries there is a first
order phase transition.
\begin{figure}
\includegraphics[width=0.45\textwidth]{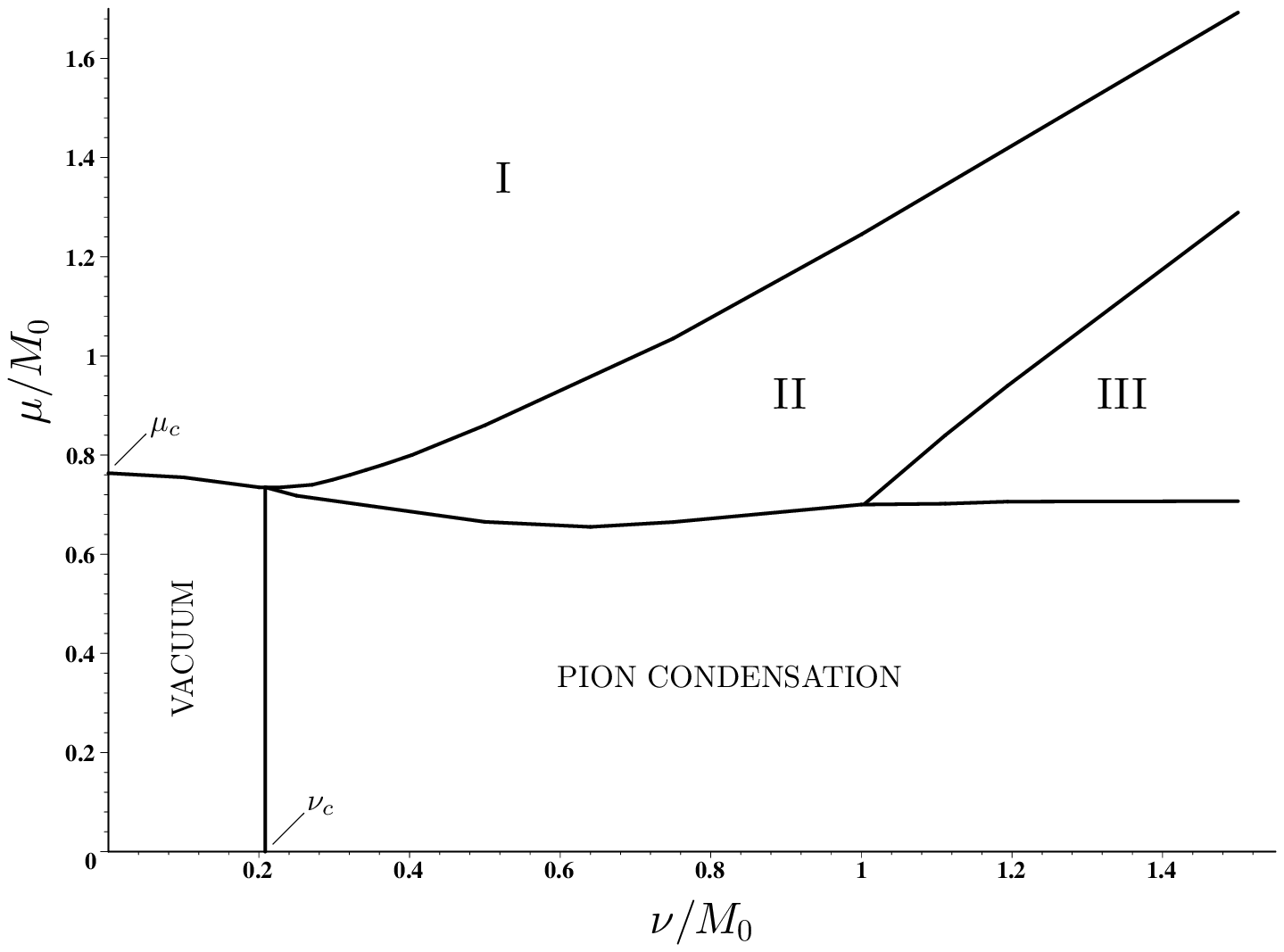}
\hfill
\includegraphics[width=0.45\textwidth]{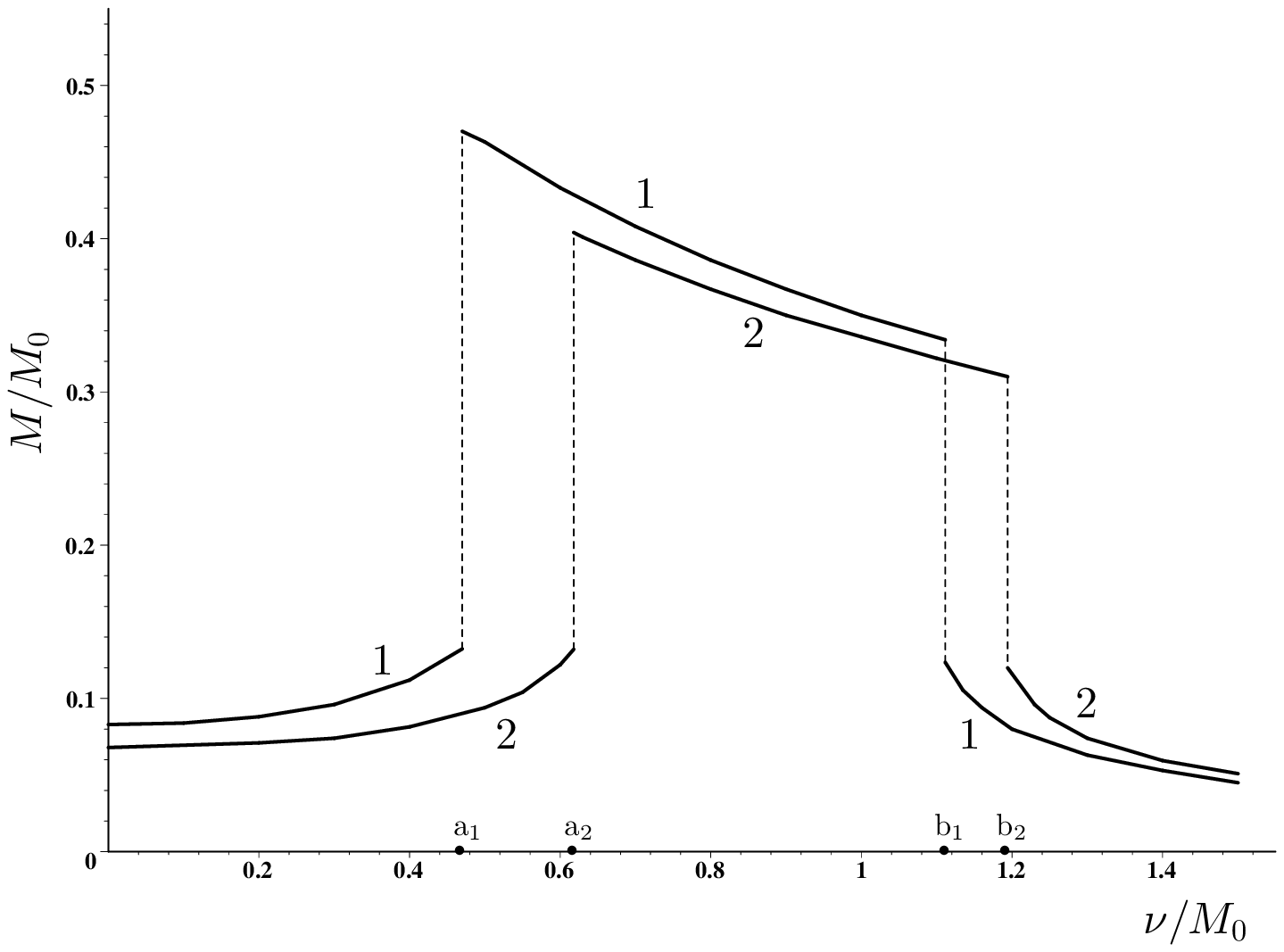}\\
\parbox[t]{0.45\textwidth}{
\caption{Phase portrait of the model in terms of $\mu$ and
$\nu\equiv\mu_I/2$ at $\alpha=\alpha_0\approx 0.17$. Here
$\nu_c/M_0\approx 0.21$, $\mu_c/M_0\approx 0.76$. All lines of the
figure are curves of first order phase transitions, except the
boundary between the vacuum and PC phase. In the normal quark matter
phases I, II and III pion condensation is absent (their properties
are discussed in the text).} } \hfill
\parbox[t]{0.45\textwidth}{
\caption{The behavior of the gap M vs $\nu$ in the phases I, II and
III at $\mu/M_0=0.84$ (curve 1) and $\mu/M_0=0.94$ (curve 2)
at $\alpha=\alpha_0\approx 0.17$. Here $a_1\approx 0.47$,
$a_2\approx 0.62$, $b_1\approx 1.11$, $b_2\approx 1.19$.}}
\end{figure}

Now, let us consider the quark number density $n_q$ (\ref{024}) as
well as the isospin density $n_I$ (\ref{025}) inside each phase of
the model. It is easy to see that for the vacuum phase these
quantities are zero, thus justifying the name of these phase. Then,
since the gaps $\Delta$ and $M$ do not depend on $\mu$ inside the
pion condensed phase and the relations $E_\Delta^\pm>\mu$ are true
here, one can conclude that $n_q\equiv 0$ in this phase and the
isospin density $n_I$ vs $\nu$ in the PC phase is presented in Fig.
4. For the normal quark matter phases I, II and III we have $\Delta
=0$, so in order to obtain the expressions for $n_q$ and $n_I$ in
these phases one can use the expression (13) of the paper \cite{ek2}
for the quantity $\Omega_{\mu,\nu}(M,\Delta =0)$. As a result, in
the phases I, II and III we have:
\begin{eqnarray}
n_q
=\frac{\theta(\mu+\nu-M)}{\pi}\sqrt{(\mu+\nu)^2-M^2}&&\!\!\!\!\!\!+
\frac{\theta(|
\mu-\nu|-M)}{\pi}\sqrt{(\mu-\nu)^2-M^2}~\mbox{sign}(\mu-\nu),
\label{29}\\
n_I=\frac{\theta(\mu+\nu-M)}{\pi}\sqrt{(\mu+\nu)^2-M^2}
&&\!\!\!\!\!\!-\frac{\theta(|
\mu-\nu|-M)}{\pi}\sqrt{(\mu-\nu)^2-M^2}~\mbox{sign}(\mu-\nu).
\label{30}
\end{eqnarray}
Then, using the values of the gap $M$ presented in Fig. 6, one can
find the corresponding values of densities $n_q$ and $n_I$ shown in
the curves of Fig. 7. It is clear from this figure that inside the
II-phase $n_q\equiv n_I$. Since $n_q=n_u+n_d$ and $2n_I=n_u-n_d$,
where $n_u$, $n_d$ are the densities of up and down quarks,
correspondingly, it is clear from the above mentioned constraint
that in the phase II the relation $n_u\equiv -3n_d$ is valid.

Up to now we have studied thermodynamic properties of the model
phases. Now the consideration of their dynamical peculiarities are
in order. The first point we would like to discuss here is the
spectrum of quasiparticles. In condensed matter physics they are
simply the one-fermion excitations of the corresponding ground state.
Recall, in the most general case the energy spectrum of $u$-, $d$-,
$\bar u$-, $\bar d$-quasiparticles (quarks) are given in
(\ref{E6}) (see Appendix \ref{ApE}). It is clear from this formula
that in the vacuum phase, where $\Delta =0$ and $M\approx 1.04M_0$,
the energy which is needed for a creation of the $u$- and
$d$-quasiparticles is always greater than zero. Hence, both $u$- and
$d$-quarks are the gapped excitations of the vacuum phase.
The similar property of a ground state is valid for the PC phase of
the model, where also a finite amount of energy is needed to create
up and/or down quarks. Due to this reason, the name gapped phases are
usually used in these cases.

However, in the case with normal quark matter phases I, II and III
the situation is opposite. Indeed, it is easy to check that in the
phase I both $u$- and $d$-quasiparticles are gapless. It
means that there are no energy costs to create these quarks, i.e.
there exist  space momenta $p_1^\star$ and $p_1^{\star\star}$ such
that $p_{0u}(p_1^\star)=0$ and $p_{0d}(p_1^{\star\star})=0$, where
$p_{0u}(p_1)$ and $p_{0d}(p_1)$ are the energies given in (\ref{E6})
of corresponding quasiparticles. (For example, the point
$(\nu=0.2M_0,\mu=0.84M_0)$ lies in the phase I with $M\approx
0.088M_0$, $\Delta =0$. Then it is easy to find from (\ref{E6}) that
$p_1^\star\approx 1.04M_0$ and $p_1^{\star\star}\approx 0.63M_0$.)
In contrast, in the phases II and III only $u$-quasiparticles are
gapless, but $d$-quarks are gapped. Note, some dynamical effects in
dense matter such as transport phenomena (e.g., conductivities etc)
depend essentially on the fact whether or not gapless excitations of
the medium are possible. Hence, these effects can occur in a
qualitatively different way in the phase I on one hand, and in the
phases II and III, on the other hand.

Finally, it is necessary to remark that the spectrum of mesonic
excitations is also has a sharp phase dependence. In particular, in
Fig. 8 the behavior of the $\pi^0$-meson mass in the phase II is
depicted at two different values of $\mu$. It turns out that in this
phase the $\pi^0$-meson is a stable particle (at least with respect
to strong interactions). However, in the neighboring phases I and
III it is no longer a stable particle but a resonance. This fact as
well as other peculiarities of the meson spectrum in different
phases of the model is the subject of our consideration in the next
section.
\begin{figure}
\includegraphics[width=0.45\textwidth]{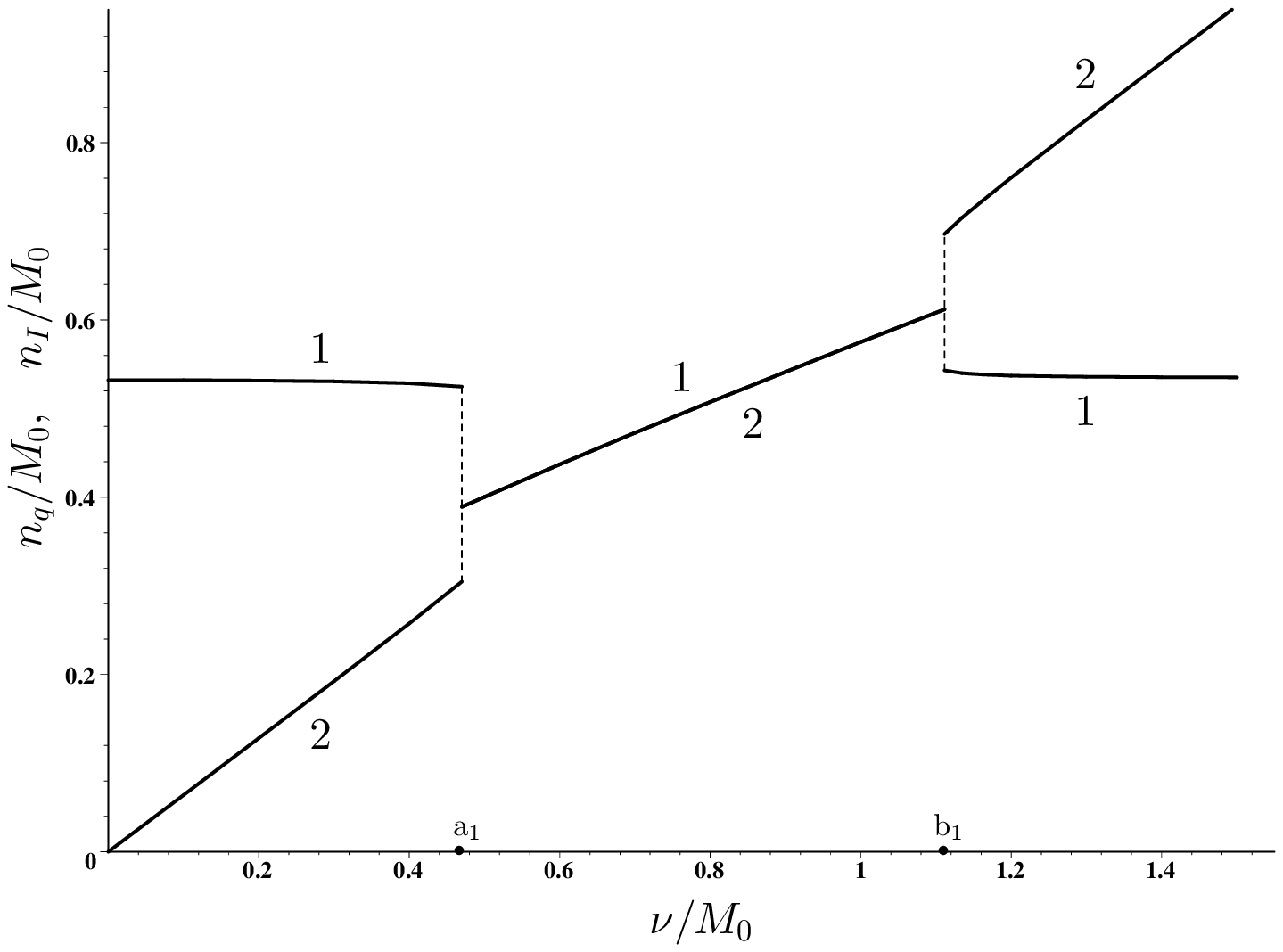}
\hfill
\includegraphics[width=0.45\textwidth]{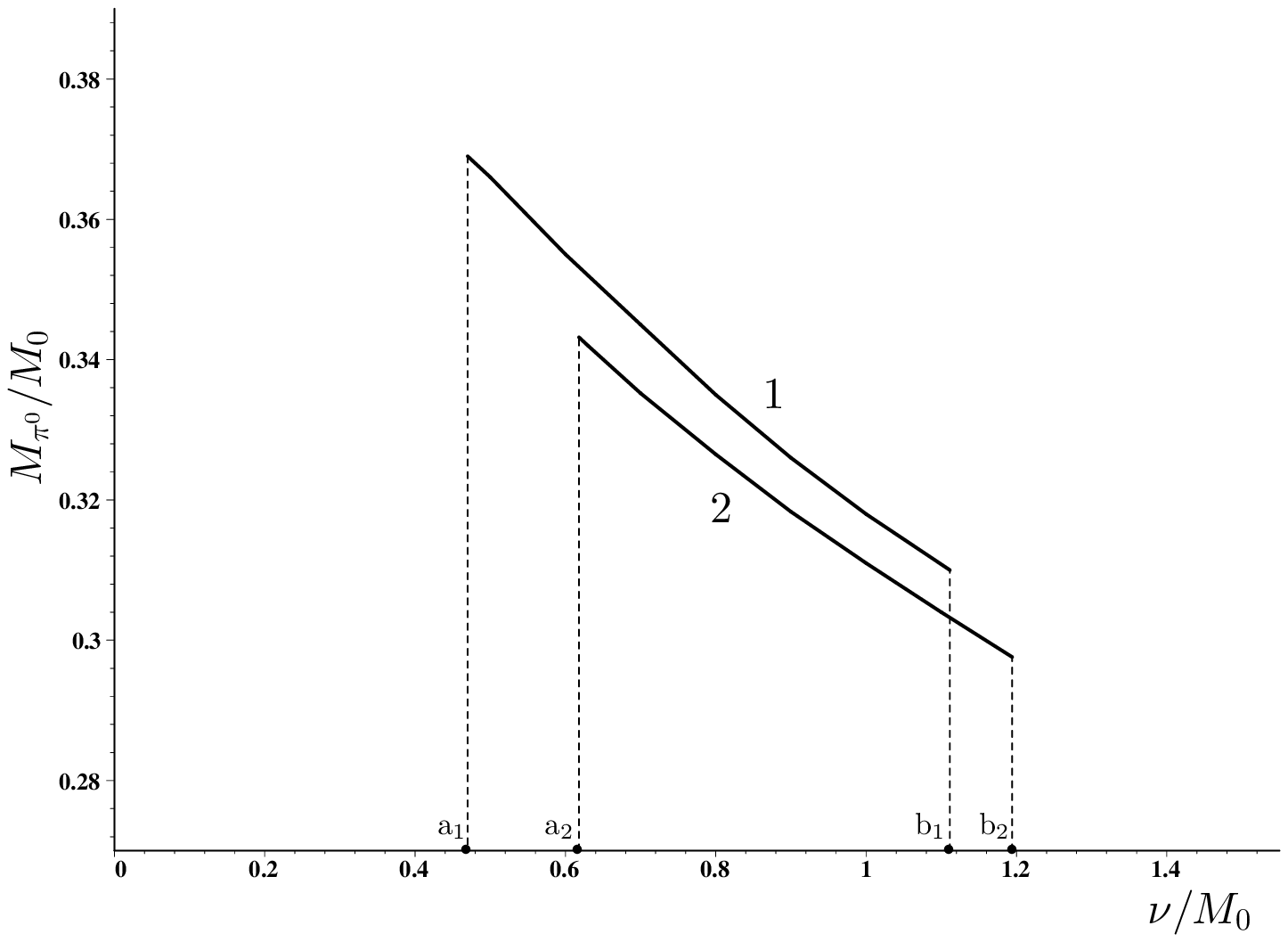}\\
\parbox[t]{0.45\textwidth}{
 \caption{Quark number density $n_q$ (curve 1) and isospin
 density $n_I$ (curve 2) vs $\nu$ at $\mu/M_0= 0.84$. Here
 $a_1\approx 0.47$, $b_1\approx 1.11$.}  }
\hfill
\parbox[t]{0.45\textwidth}{
\caption{The behavior of the $\pi^0$-meson mass $M_{\pi^0}$ vs $\nu$
in the phase II at $\mu/M_0=0.84$ (curve 1) and $\mu/M_0=0.94$
(curve 2). The values $a_i$ and $b_j$ are given in Fig. 6.}}
\end{figure}

\section{Meson masses in different phases}

As was noted in section \ref{effaction} (see the text after
(\ref{A11})), the effective action (\ref{A11}) can be used for
obtaining meson masses in different phases of the model. For this
purpose, one should find from the outset all two-point 1PI Green
functions (\ref{A12}) of meson fields. These 1PI Green functions are
the matrix elements of the 4$\times$4 meson matrix $\Gamma (x-y)$.
Then it is necessary to get the Fourier transformation
$\overline{\Gamma} (p)$ of the meson matrix $\Gamma (x-y)$ and find
its determinant in the rest frame, where the two-component
energy-momentum vector $p$ has the form $p=(p_0,0)$. The equation
\begin{eqnarray}
\mbox{det}\overline{\Gamma} (p_0)=0\label{31}
\end{eqnarray}
has in the plane of the variable $p_0^2$ four (real- or
complex-valued) solutions, one of them is the mass squared of the
scalar $\sigma$-meson, whereas the other three solutions give the
mass squared  of the pseudoscalar $\pi$-mesons.

Detailed investigations of the meson matrix $\overline{\Gamma}
(p_0)$ show that its matrix elements of the form
$\overline{\Gamma}_{\pi_3X} (p_0)$ or $\overline{\Gamma}_{X\pi_3}
(p_0)$, where $X=\sigma,\pi_1,\pi_2$, are equal to zero in all
phases of the model, i.e. the matrix $\overline{\Gamma} (p_0)$ is a
reducible one. This means that the neutral pseudoscalar meson,
$\pi^0\equiv\pi_3$, does not mix with the other mesons,
$\pi^\pm\equiv (\pi_1\pm\mathrm{i}\pi_2)/\sqrt{2}$ or $\sigma$. As a
result, one root of the equation (\ref{31}) can be found through the
equation $\overline{\Gamma}_{\pi_3\pi_3} (p_0)=0$, which supplies us
with the mass squared of the $\pi^0$-meson in different phases of
the model. The other three meson masses are the zeros of the
determinant of the reduced meson matrix, whose matrix elements are
two point 1PI Green functions of the fields $\sigma$, $\pi_1$, and
$\pi_2$.

\subsection{The mass of $\pi^0$-meson}

The corresponding two-point 1PI Green
function looks like:
\begin{eqnarray}
\Gamma_{\pi^0\pi^0} (z)\equiv -\frac{\delta^2{\cal S}^{(2)}_{\rm
eff}}{\delta\pi_3(y)\delta\pi_3(x)}&=&\frac{\delta(z)}{2G}+
\mathrm{i}{\rm Tr}_{s}
\big\{S_{11}(z)\gamma^5S_{11}(-z)\gamma^5+S_{22}(z)\gamma^5
S_{22}(-z)\gamma^5\nonumber\\&-&
S_{12}(z)\gamma^5S_{21}(-z)\gamma^5-S_{21}(z)
\gamma^5S_{12}(-z)\gamma^5\big\},\label{32}
\end{eqnarray}
where $z=x-y$, and the matrix elements $S_{ij}(z)$ of the quark
propagator are presented in (\ref{E4}). Note, the expression
(\ref{32}) is valid for all phases of the model. Now, let us
consider it in each phase.

\subsubsection{Vacuum and normal quark matter phase I: The case
$\nu=0$, $\mu\ge 0$}\label{mesmas}

To illustrate the technique, which was elaborated in details in the
framework of NJL-models with the color superconductivity phenomenon
\cite{eky}, we start from the most simple case of $\nu=0$
corresponding to the vacuum and the phase I only (see section
\ref{mu} and Fig. 5). Since for these phases $\Delta=0$, the last
two terms in (\ref{32}), proportional to $S_{12}(z)$, vanish. The
corresponding Fourier transformation of the expression (\ref{32})
now looks like:
\begin{eqnarray}
&&\overline{\Gamma}_{\pi^0\pi^0}(p)=\frac{1}{2G}+\mathrm{i} {\rm
Tr}_{s}\int\frac{d^2k}{(2\pi)^2}
\left\{\overline{S}_{11}(p+k)\gamma^5\overline{S}_{11}(k)\gamma^5+
\overline{S}_{22}(p+k)\gamma^5\overline{S}_{22}(k)\gamma^5\right\},
\label{D3}
\end{eqnarray}
where the Fourier transformations $\overline{S}_{ij}(p)$ can be
easily determined from (\ref{E4}). Using in (\ref{D3}) the rest
frame system, where $p=(p_0,0)$, and calculating the trace over
spinor indices, we have
\begin{eqnarray}
&&\overline{\Gamma}_{\pi^0\pi^0}(p_0)=\frac{1}{2G}+4\mathrm{i}
\int\frac{d^2k}{(2\pi)^2}\frac{E^2-(k_0+\mu)(p_0+k_0+\mu)}
{((k_0+\mu)^2-E^2)[(k_0+p_0+\mu)^2-E^2]}, \label{D4}
\end{eqnarray}
where $E=\sqrt{k_1^2+M^2}$, and the dynamical quark mass $M$ is
given by the value of the $M$-coordinate of the GMP of the
thermodynamic potential. As was noted in Appendix \ref{ApE}, in
(\ref{D4}) $k_0$ and $(k_0+p_0)$ are correspondingly the shorthand
notations for $k_0+\mathrm{i}\varepsilon\cdot {\rm sign}(k_0)$ and
$(k_0+p_0)+\mathrm{i}\varepsilon\cdot {\rm sign}(k_0+p_0)$, where
$\varepsilon\to 0_+$. The $k_0$-integration in (\ref{D4}) is
performed along the real axis in the complex $k_0$-plane. We will
close this contour by an infinite arc in the upper half of the
complex $k_0$-plane. Taking into account the above-mentioned rule
for the $k_0$-integration, we have inside the obtained closed
contour of the integral in (\ref{D4}) four poles of the integrand
which are located in the following points:
\begin{eqnarray}
(k_0)_1=-E-\mu+\mathrm{i}\varepsilon\theta
(\mu+E),~~~&&~~~(k_0)_2=E-\mu+\mathrm{i}\varepsilon\theta
(\mu-E),\nonumber\\(k_0)_3=-E-\mu-p_0+\mathrm{i}\varepsilon\theta
(\mu+E),~~~&&~~~(k_0)_4=E-\mu-p_0+\mathrm{i}\varepsilon\theta
(\mu-E). \label{D5}
\end{eqnarray}
Since $\mu\ge 0$ and $E\ge 0$, the sum of the corresponding residues
of the integrand function in these poles results in the following
$k_1$-integration in (\ref{D4}):
\begin{eqnarray}
&&\overline{\Gamma}_{\pi^0\pi^0}(p_0)=\frac{1}{2G}+8
\int_{-\infty}^{\infty}\frac{dk_1}{2
\pi}\frac{E\theta(E-\mu)}{p_0^2-4E^2}. \label{D6}
\end{eqnarray}
To renormalize the expression (\ref{D6}) we use the gap equation
(\ref{013}) at $\Delta=\nu=0$:
\begin{eqnarray}
\frac{1}{2G}=\frac{m}{2M}+2\int_{-\infty}^{\infty}\frac{dk_1}{2\pi}
\frac{\theta(E-\mu)}{E}, \label{D7}
\end{eqnarray}
where we took into account that $m_0/G\equiv m$. Substituting
(\ref{D7}) into (\ref{D6}) and using the relation $m\equiv \alpha
M_0/\pi$, we have
\begin{eqnarray}
&&\overline{\Gamma}_{\pi^0\pi^0}(p_0)= \frac{\alpha M_0}{2\pi M}-
2p_0^2\int_{0}^{\infty}\frac{dk_1}{\pi}\frac{\theta(E-\mu)}{E(4E^2-
p_0^2)}. \label{D8}
\end{eqnarray}
Note that the quantity $\overline{\Gamma}_{\pi^0\pi^0}(p_0)$ is a
multi-valued function of the variable $p_0^2$ which is analytic on
some complex Riemann manifold described by several sheets. The
expression in the right hand side of (\ref{D8}) defines
$\overline{\Gamma}_{\pi^0\pi^0}(p_0)$ just on the first physical
sheet only, which is the whole complex $p_0^2$-plane, except for the
cut $p_0^2>4M^2$ along the real axis.

Recall that the mass squared $M_{\pi^0}^2$ of $\pi^0$-mesons is the
zero of this 1PI Green function vs $p_0^2$. The zero should lie
either on the real axis in the first sheet of the  $p_0^2$-plane (in
this case it corresponds to a stable particle with real value of
$M_{\pi^0}^2$ such that $0\le M_{\pi^0}^2\le 4M^2$) or in the second
sheet, corresponding to a resonance. Since at  $\nu=0$ a mass
splitting between $\pi$-mesons is absent (see also the remark in the
footnote \ref{vacrem}), throughout the section we use the notation
$M_\pi$ both for the $\pi^0$- as well as for the $\pi^\pm$-meson
mass.

It is clear from Fig. 1 (see also section \ref{mu}) that in the
vacuum phase at $\nu=0$ the relation $\mu<\mu_c<M$ is valid for
arbitrary $\alpha$-values, so the theta-function in (\ref{D8}) is
equal to unity. As a result, we see that in the vacuum the
$\pi$-meson mass satisfies the following equation:
\begin{eqnarray}
&&\frac{\alpha M_0}{2M}=
2M_\pi^2\int_{0}^{\infty}\frac{dk_1}{\sqrt{k_1^2+M^2}(4k_1^2+4M^2-
M_\pi^2)}. \label{C11}
\end{eqnarray}
Supposing that the quantity $M$ in (\ref{C11}) is just the gap
depicted in Fig. 1 as the curve 1, one can solve numerically this
equation with respect to the variable $M_\pi$. It turns out that the
solution $M_\pi$ lies in the first sheet of the Riemann manifold and
hence obeys the relation $M_\pi^2<4M^2$. The quantity $M_\pi$ vs
$\alpha$ is shown in Fig. 1 as the curve 2.

In contrast, in the case of the phase I at $\nu=0$ the corresponding
1PI Green function (\ref{D8}) does not has zeros in the first
Riemann sheet of the variable $p_0^2$, i.e. there are no stable (at
least with respect to strong interactions) $\pi$-mesonic excitations
of the phase I ground state. In this phase all $\pi$-mesons are
resonances.

\subsubsection{Vacuum and normal quark matter phases I, II and III:
The case $\nu\ne 0$, $\mu\ne 0$}

Technically this is a more complicated case, but the main ideas of
calculations do not change. So, omitting technical details, one can
obtain the following expression for the two-point 1PI Green function
of $\pi^0$-mesons in the rest frame:
\begin{eqnarray}
&&\overline{\Gamma}_{\pi^0\pi^0}(p_0)= \frac{\alpha M_0}{2\pi M}-
p_0^2\int_{0}^{\infty}\frac{dk_1}{\pi}\frac{1}{E(4E^2- p_0^2)}\big
[\theta(E+\nu-\mu)+\mbox{sign}(E-\nu)\theta(|E-\nu|-\mu)\big ].
\label{D10}
\end{eqnarray}
It is also a multi-valued function of the variable $p_0^2$ which is
analytical on the same Riemann manifold, where the Green function
(\ref{D8}) is defined. On the first Riemann sheet and at real values
of $p_0^2$ such that $0\le p_0^2<4M^2$ it looks like:
\begin{eqnarray}
&&\overline{\Gamma}_{\pi^0\pi^0}(p_0)= \frac{\alpha M_0}{2\pi M}-
\frac{2p_0}{\sqrt{4M^2-p_0^2}}\arctan
\frac{p_0}{\sqrt{4M^2-p_0^2}}+\frac{p_0~
\theta(\mu+\nu-M)}{\sqrt{4M^2-p_0^2}}\arctan\frac{p_0
\sqrt{(\mu+\nu)^2-M^2}}{(\mu+\nu)\sqrt{4M^2-p_0^2}}\nonumber\\
&&~~~~~~~~~~~~~~~+\frac{p_0~
\theta(|\mu-\nu|-M)}{\sqrt{4M^2-p_0^2}}\arctan\frac{p_0
\sqrt{(\mu-\nu)^2-M^2}}{|\mu-\nu|\sqrt{4M^2-p_0^2}}. \label{D11}
\end{eqnarray}
Let us, for example, again consider the case $\alpha=\alpha_0\approx
0.17$, then $M\approx 1.04 M_0$ (see Fig. 2). In this case, for
the $\mu$ and $\nu$ values taken from the vacuum phase of Fig. 5,
each theta-function in the expression (\ref{D10}) is equal to unity.
As a result, the Green function (\ref{D10}) at $\nu\ne 0$ coincides
with the $\pi^0$ Green function (\ref{D8}) at $\nu=0$. Hence, in the
vacuum phase the mass of the $\pi^0$-meson does not depend on both
$\nu$ and $\mu$. It takes the value $M_{\pi^0}\approx 0.42 M_0$
in the case of $\alpha=\alpha_0$. In the general case of arbitrary
$\alpha$-values, the $M_{\pi^0}$-mass in the vacuum phase at $\nu\ne
0$ is simply the pion mass at $\nu =0$ (see the line 2 of Fig. 1).

One can easily check that the expression (\ref{D11}) turns into zero
at some point of the interval $0< p_0<2M$ only in the case when
$(\nu,\mu)$ lies in the phase II (the corresponding value of $p_0$
is the mass of the $\pi^0$-meson). At some fixed values of $\mu$ the
behavior of $M_{\pi^0}$ vs $\nu$ is presented in Fig. 8 at
$\alpha=\alpha_0$. In contrast, in the phases I and III the
expression (\ref{D11}) has no zeros in the interval $0< p_0<2M$.
Hence, in these phases $\pi^0$ is not a stable particle, but rather
a resonance.

\subsubsection{The pion condensation phase}

Now, let us study the $\pi^0$-mass in the PC phase, where both gaps
$\Delta$ and $M$ are nonzero. To obtain a compact expression for the
two-point 1PI Green function $\overline{\Gamma}_{\pi^0\pi^0}(p_0)$,
it is again necessary to eliminate in (\ref{32}) the coupling
constant with the help of the gap equation (\ref{13}), i.e. to use
the following relation
\begin{eqnarray*}
\frac{1}{2G}=\int_{-\infty}^{\infty}\frac{dk_1}{2\pi}\Big
\{\frac{\theta(E_\Delta^+-\mu)}{E_\Delta^+}+
\frac{\theta(E_\Delta^--\mu)}{E_\Delta^-} \Big\}. \label{130}
\end{eqnarray*}
Then, after tedious but straightforward calculations which are
similar to that of section \ref{mesmas}, it is possible to find
\begin{eqnarray}
&&\overline{\Gamma}_{\pi^0\pi^0}(p_0)=(p_0^2-\mu_I^2)\int_{-\infty}^{
\infty}\frac{dk_1}{2\pi}\frac{E_\Delta^++E_\Delta^-}
{E_\Delta^-E_\Delta^+[p_0^2-(E_\Delta^++E_\Delta^-)^2]}. \label{69}
\end{eqnarray}
Clearly, the mass of $\pi^0$ in the PC phase is equal to the
isotopic chemical potential $\mu_I$ and does not depend on $\mu$.

\subsection{The masses of $\sigma$- and $\pi^\pm$-mesons}

As was noted above, to get the masses of $\sigma$- and
$\pi^\pm$-mesons, it is necessary to find the zeros (in the rest
frame with $p=(p_0,0)$) of the determinant of the reduced meson
matrix composed from two-point 1PI Green functions of these
particles. Our calculations show that the Green functions are of the
form $\overline{\Gamma}_{\sigma\pi_{1,2}}(p_0)\sim \Delta$. So, in
the vacuum as well as in the phases I, II and III there is no mixing
between $\sigma$- and $\pi_{1,2}$-fields which leads to a further
reduction of the meson matrix. Hence, to find the mass of the
$\sigma$-meson in these phases, it is sufficient to investigate the
separate equation $\overline{\Gamma}_{\sigma\sigma}(p_0)=0$. The
equation $\mbox{det}\Pi(p_0)=0$ with
\begin{eqnarray}
\Pi(p_0)\equiv\left (\begin{array}{cc}
\overline{\Gamma}_{\pi_1\pi_1}(p_0)~,&\overline{\Gamma}_{\pi_1\pi_2}(
p_0) \\
\overline{\Gamma}_{\pi_2\pi_1}(p_0)~,&\overline{\Gamma}_{\pi_2\pi_2}(
p_0)\end{array}\right ),\label{70}
\end{eqnarray}
then supplies us with the masses of $\pi^\pm$-mesons.

\subsubsection{$\sigma$-meson in vacuum and I, II, III phases}

In these phases $\Delta=0$. So, On the basis of the effective action
(\ref{A11}) and using the relation (\ref{A12}) and the methods of
the previous section \ref{mesmas}, it is possible to obtain the most
general expression for the two-point 1PI Green function of the
$\sigma$-meson both in vacuum and in the I-, II-, III phases of the
model
\begin{eqnarray}
\overline{\Gamma}_{\sigma\sigma} (p_0)&=&\frac{\alpha M_0}{2\pi M}-
(p_0^2-4M^2)\int_{0}^{\infty}\frac{dk_1}{\pi}\frac{1}{E(4E^2-
p_0^2)}\big
[\theta(E+\nu-\mu)+\mbox{sign}(E-\nu)\theta(|E-\nu|-\mu)\big ].
 \label{71}
\end{eqnarray},
where again $E=\sqrt{k_1^2+M^2}$.  Let us now suppose that the pair
of chemical potentials $(\mu,\nu)$ belongs to the vacuum phase of
Fig. 5, where, evidently, $M>\mu+\nu$. In this particular case the
expression in the square brackets of (\ref{71}) is equal to 2, so
\begin{eqnarray}
&&\overline{\Gamma}^{vac}_{\sigma\sigma} (p_0)= \frac{\alpha
M_0}{2\pi M}-
2(p_0^2-4M^2)\int_{0}^{\infty}\frac{dk_1}{\pi}\frac{1}{E(4E^2-
p_0^2)}. \label{72}
\end{eqnarray}
It follows from (\ref{72}) that in the chiral limit, when $\alpha=0$
and $M\ne 0$, the $\sigma$-meson is a stable particle with mass
equal to $2M$. However, at arbitrary small $\alpha>0$ the zero of
the Green function (\ref{72}), located at the point $p_0^2=4M^2$ of
the first Riemann sheet at $\alpha=0$, shifts to the second Riemann
sheet, signalling thus that in the vacuum phase of the massive GN
model the $\sigma$-meson is a resonance. It is quite reasonable that
at small values of $\alpha$ the mass of this resonance is near $2M$.

Now remark that for values of $\mu$ and $\nu$ from the regions I, II
or III of Fig. 5 the square brackets of the integrand in (\ref{71}).
cannot be negative. As a result, for all real values of $p_0^2$ such
that $0<p_0^2<4M^2$ the Green function
$\overline{\Gamma}_{\sigma\sigma} (p_0)$ is a positive quantity,
i.e. it cannot become zero. Thus, in the phases I, II and III of the
model the $\sigma$-meson is also a resonance. \footnote{Strictly
speaking, the found resonance character of $\sigma$ is here
associated to the existence of $q \bar q$-thresholds enabling the
meson decay into a (non-observable) $q \bar q$-pair. Clearly, in
order to model e.g. the confinement properties of "more realistic"
two-dimensional QCD \cite{hooft}, one should consider a more
sophisticated GN-model incorporating some suitable prescription for
quark confinement (see, e.g., \cite{EFR}). Within such a model one
could then treat the $\sigma$-decay into observable pions, $\sigma
\to\pi\pi$, which is, however, outside the scope of this paper.}

\subsubsection{$\pi^\pm$-mesons in vacuum and I, II, III phases}
\label{mass1}

The squared masses of $\pi^\pm$-mesons in these phases are given by
the zeros of the equation $\mbox{det}\Pi(p_0)=0$ in the
$p_0^2$-plane, where $\Pi(p_0)$ is the matrix (\ref{70}). To find
its matrix elements, it is convenient to use in the effective action
(\ref{A11}) the new fields
$\pi^\pm(x)=(\pi_1(x)\pm\mathrm{i}\pi_2(x))/\sqrt{2}$ instead of the
old ones, $\pi_{1,2}(x)$. Then, it is natural to define the
corresponding Green functions $\Gamma_{\pi^+\pi^-}(x-y)$ etc, where
\begin{eqnarray*}
&& \Gamma_{\pi^+\pi^-}(x-y)=-\frac{\delta^2{\cal S}^{(2)}_{\rm
eff}}{\delta\pi^-(y)\delta\pi^+(x)}
\label{73}
\end{eqnarray*}
etc. \footnote{In the phases with zero gap $\Delta$ the
Green functions of the form $\Gamma_{\pi^+\pi^+}(x-y)$ and
$\Gamma_{\pi^-\pi^-}(x-y)$ vanish.}
The Fourier transformations of these Green functions are connected
with the matrix elements of the matrix $\Pi(p_0)$ (\ref{70}) by the
relations
\begin{eqnarray}
&&\overline{\Gamma}_{\pi_1\pi_1}(p_0)=\overline{\Gamma}_{\pi_2\pi_2}(
p_0)=\frac{1}{2}\left
[\overline{\Gamma}_{\pi^+\pi^-}(p_0)+\overline{\Gamma}_{\pi^-\pi^+}(p
_0)\right ],\nonumber\\
&&\overline{\Gamma}_{\pi_1\pi_2}(p_0)=-\overline{\Gamma}_{\pi_2\pi_1}
(p_0)=\frac{\mathrm{i}}{2}\left
[\overline{\Gamma}_{\pi^+\pi^-}(p_0)-\overline{\Gamma}_{\pi^-\pi^+}(p
_0)\right ].\label{74}
\end{eqnarray}
Then, the determinant of the matrix (\ref{70}) looks like:
\begin{eqnarray}
\mbox{det}\Pi(p_0)=\overline{\Gamma}_{\pi^+\pi^-}(p_0)\cdot
\overline{\Gamma}_{\pi^-\pi^+}(p_0).\label{75}
\end{eqnarray}
Our straightforward analytical calculations show that
\begin{eqnarray}
\overline{\Gamma}_{\pi^+\pi^-}(p_0)=\overline{\Gamma}_{\pi^0\pi^0}(p_
0+\mu_I),~~~
\overline{\Gamma}_{\pi^-\pi^+}(p_0)=\overline{\Gamma}_{\pi^0\pi^0}(
\mu_I-p_0),\label{76}
\end{eqnarray}
where $\overline{\Gamma}_{\pi^0\pi^0}$ is the 1PI Green function of
the $\pi^0$-meson, presented in (\ref{D10}). Now suppose that at
$p_0^2=M_{\pi^0}^2$ the Green function of the $\pi^0$-meson turns
into zero, when the chemical potentials $(\mu,\mu_I)$ are fixed at
some values in the vacuum phase or the I-, II-, III phases. Then, on
the basis of the relations (\ref{76}) it is clear that
$\overline{\Gamma}_{\pi^+\pi^-}(p_0)\sim
((p_0+\mu_I)^2-M_{\pi^0}^2)$ and
$\overline{\Gamma}_{\pi^-\pi^+}(p_0)\sim
((\mu_I-p_0)^2-M_{\pi^0}^2)$. As a result, we see that
\begin{eqnarray}
\mbox{det}\Pi(p_0)\sim [(p_0+\mu_I)^2-M_{\pi^0}^2]\cdot
[(\mu_I-p_0)^2-M_{\pi^0}^2]\equiv [p_0^2-(M_{\pi^0}-\mu_I)^2]\cdot
[p_0^2-(M_{\pi^0}+\mu_I)^2].\label{77}
\end{eqnarray}
Hence, the zeros of the determinant (\ref{77}), i.e.
the quantities $M_{\pi^+}^2=(M_{\pi^0}-\mu_I)^2$ and
$M_{\pi^-}^2=(M_{\pi^0}+\mu_I)^2$, can be identified with the mass
squared of $\pi^\pm$-mesons.

\subsubsection{$\sigma$- and $\pi^\pm$-mesons in the pion
condensation phase}

As noted at the beginning of the present section, there arises a
mixing between $\sigma$ and $\pi_{1,2}$ fields in the PC phase of
the massive GN model. Thus, to define the mesonic mass spectrum one
should find all the zeros of the determinant of the meson matrix,
composed of corresponding two-point 1PI Green functions of the form
(\ref{A12}). We have found an exact analytical expressions for these
Green functions and have shown that the determinant has a zero in
the point $p_0^2=0$. (In order not to overload the paper with rather
cumbersome formulae, we do not present here the expressions for
these Green functions.) It means that in the PC phase there is a
massless bosonic excitation. It can be treated as a Goldstone boson
which is a consequence of the spontaneous breaking of the isospin
$U_{I_3}(1)$ symmetry in the PC phase.

It turns out that further information about mesons in the PC phase
can be found in the chiral limit, i.e. at $m_0=0$. Indeed, in this
case the Green functions of the form
$\overline{\Gamma}_{\sigma\pi_{1,2}}(p_0)$ are identically equal to
zero, so that the $\sigma$-meson does not mix with
$\pi_{1,2}$-fields. Moreover, it is possible to show that in the
massless GN model the Green function
$\overline{\Gamma}_{\sigma\sigma}(p_0)$ coincides in
the PC phase with the Green function
$\overline{\Gamma}_{\pi^0\pi^0}(p_0)$ (see (\ref{69})). Due to this
relation we conclude that $M_\sigma=M_{\pi^0}=\mu_I$ in the PC phase
of the massless GN model.

\section{Summary and conclusions}

Recent investigations of the phase diagram of isotopically
asymmetric dense quark matter in terms of NJL models show that their
pion condensation content is not yet fully understood. Indeed, the
number of the charged pion condensation phases of the phase diagram
depends strictly on the parameter set of the NJL model. It
means that for different values of the coupling constant, cutoff
parameter, bare quark mass etc just the same NJL model predicts
different numbers of pion condensation phases of quark matter both
with or without an electric neutrality constraint (see, e.g.,
\cite{ek,andersen}). Thus, to obtain a more objective information
about the pion condensation phenomenon of dense quark matter, it is
important to invoke alternative approaches. One of them, which
qualitatively quite successfully imitates some of the QCD properties
(see also the Introduction), is based on the consideration of this
phenomenon in the framework of asymptotically free (1+1)-dimensional
GN models in the leading order of the large $N_c$-technique.

In the present paper we have studied the phase structure of the
massive GN model (1) in terms of  quark number ($\mu$)- as well as
isospin ($\mu_I$) chemical potentials in the limit $N_c\to\infty$
(for simplicity, the temperature has been taken to be zero). After
renormalization (comp. section \ref{omega}), this model contains two
free parameters: $M_0$--the dynamical quark mass in the vacuum of
the corresponding massless GN model and the
renormalization--invariant quark mass $m=m_0/G\equiv\alpha M_0/\pi$
(see also the remark in footnote \ref{foot1}). In our considerations
we often put $\alpha=\alpha_0\approx 0.17$ in order to have the same
relation between the dynamical quark mass $M$ and the $\pi$-meson
mass $M_\pi$ in vacuum, i.e. $M/M_\pi =5/2$, as used in some other
NJL model parameterizations \cite{eky}. Just at $\alpha=\alpha_0$
the phase portrait of the model is presented in Fig. 5 in terms of
$\mu$ and $\nu=\mu_I/2$.

Firstly, we have found that at $T=0$ the charged pion condensation
phase of the GN model is realized inside the (noncompact) chemical
potential region $\mu_I>M_{\pi^0}$, where $\mu$ is not greater than
$M_0/\sqrt{2}$ and $M_{\pi^0}$ is the vacuum mass of the
$\pi^0$-meson. In this phase the isospin $U_{I_3}(1)$ symmetry is
spontaneously broken down and a massless Goldstone bosonic
excitation of the ground state appears. Moreover, we have shown that
the mass of the $\pi^0$-meson in the PC phase is equal to the
isospin chemical potential $\mu_I$.  All one-quark excitations are
found to be gapped particles in this phase. As a result, the quark
number density $n_q$ is equal to zero in the PC phase. \footnote{In
the gapped phases, PC or vacuum phases, the relations
$E_\Delta^\pm>\mu$ are valid. Then, using (\ref{024}), it is clear
that $n_q\equiv 0$ in these phases.} The same properties of the PC
phase is predicted in the framework of some NJL model
parameterizations (see, e.g., in \cite{ek,andersen}). In contrast,
in the NJL phase diagram the pion condensation phases occupy a
compact region and for some parametrization schemes the gapless pion
condensation might occur \cite{frank,ek,jin,andersen}.

Secondly, at rather large values of the quark number chemical
potential $\mu$  we have found a rather rich variety of normal quark
matter phases I, II, and III (see Fig. 5), in which the quark number
density $n_q$ does not vanish (see Fig. 7). In particular, it turns
out that in phase I both $u$- and $d$-quarks are gapless
quasiparticles. Contrary, in phases II and III only $u$-quarks are
gapless, whereas $d$-quarks are gapped. By this reason, dynamical
effects in transport phenomena for dense quark matter (e.g.,
conductivities etc) can occur in a qualitatively different way in
the phases I and II, III. We have studied also the $\pi$-meson mass
spectrum of these phases and found that in the phase I and III the
$\pi$-mesons are resonances. However, the phase II is the so-called
"stability island" for $\pi$-mesons. Indeed, as it was shown by our
numerical calculations, the $\pi^0$-meson is a stable excitation of
the ground state of this phase. Its mass vs $\nu$ is depicted in
Fig. 8. The $\pi^\pm$-mesons are also stable in this phase, but
their masses are $M_{\pi^\pm}=|M_{\pi^0}\mp\mu_I|$ (see section
\ref{mass1}). (The same relation between $\pi^0$- and
$\pi^\pm$-meson masses is also valid inside the vacuum phase of Fig.
5.)

In conclusion, by using the above rather simple approach to the GN
phase diagram, we have found a variety of phases with rather rich
dynamical contents. A related interesting issue could be the
extension of these investigation to inhomogeneous condensates
\cite{dunne}. We hope that our investigation of the phase diagram of
the massive GN model will shed some new light on the phase structure
of QCD at nonzero baryonic and isotopic densities. Obviously, a more
realistic imitation of the QCD phase diagram requires to include
also a nonzero temperature as well as a suitable confinement
prescription for quark propagators \cite{EFR}.

\section*{Acknowledgments}

One of us (K.G.K.) is grateful to Prof. M. Mueller-Preussker and his
colleagues for the kind hospitality at the Institute of Physics of
the Humboldt-University and to {\it Deutscher Akademischer
Austauschdienst} (DAAD) for financial support.

\appendix

\section{Traces of operators and their products}
\label{ApB}

Let $\hat A,\hat B,...$ are some operators in the Hilbert space
$\mathbf H$ of functions $f(x)$ depending on two real variables,
$x\equiv (x^0,x^1)$. In the coordinate representation their matrix
elements are $A(x,y), B(x,y) ,...$, correspondingly, so that
 $$(\hat A f)(x)\equiv \int d^2yA(x,y)f(y),~~(\hat A\cdot \hat
 B)(x,y)\equiv \int
 d^2zA(x,z)B(z,y), ~~\mbox{etc}$$
 By definition,
 \begin{eqnarray}
{\rm Tr}\hat A\equiv\int d^2xA(x,x),~~{\rm Tr}(\hat A\cdot\hat
B)\equiv\int d^2xd^2yA(x,y)B(y,x),~~\mbox{etc}.
\label{B0}
\end{eqnarray}
Each function $f(x)\in \mathbf H$ can be considered as an operator
$\hat f$, acting in this space, with matrix elements $f(x)\delta
(x-y)$, where $\delta (x-y)$ is the two-dimensional Dirac
delta-function. As a result, one can formally consider the trace of
functions, their products as well as the traces of more complicated
expressions, such as the products of operators and functions.
Indeed, using the definition (\ref{B0}) we have
\begin{eqnarray}
{\rm Tr}f\equiv {\rm Tr}\hat f =\delta(0)\int
d^2xf(x);~~~{\rm
Tr}(f_1 f_2)&\equiv& {\rm Tr}(\hat f_1\cdot\hat f_2)=\int
d^2xd^2yf_1(x)\delta (x-y)f_2(y)\delta (y-x)\nonumber\\
=\delta(0)\int d^2xf_1(x)f_2(x);~~{\rm Tr}\{\hat A f\}\equiv {\rm
Tr}\{\hat A\cdot\hat f\}&=&\int d^2xd^2y A(x,y)f(y)\delta(y-x)=\int
d^2x A(x,x)f(x);\nonumber\\
{\rm Tr}\{\hat A f_1\hat B f_2\}\equiv{\rm Tr}\{\hat
A\cdot\hat f_1\cdot\hat B\cdot\hat f_2\}&=&\int d^2xd^2yd^2vd^2u
A(x,v)f_1(v)\delta(v-y)B(y,u)f_2(u)\delta(u-x)\nonumber\\
&=&\int d^2ud^2v A(u,v)f_1(v)B(v,u)f_2(u). \label{B1}
\end{eqnarray}
In particular, it follows from (\ref{B1}) that
\begin{eqnarray}
\frac{\delta{\rm Tr}\{\hat A f\}}{\delta f(x)}=A(x,x);~~~~
\frac{\delta^2{\rm Tr}\{\hat A f_1\hat
Bf_2\}}{\delta f_1(x)\delta f_2(y)}=
A(y,x)B(x,y).
\label{B2}
\end{eqnarray}
Now suppose that $A(x,y)\equiv A(x-y)$, $B(x,y)\equiv B(x-y)$, i.e.
that $\hat A,\hat B$ are translationally invariant operators. Then
introducing the Fourier transformations of their matrix elements,
i.e.
\begin{eqnarray}
\overline{A}(p)=\int d^2z
A(z)e^{ipz},
~~~~~~~ A(z)=\int\frac{d^2p}{(2\pi)^2}
\overline{A}(p)e^{-ipz},~~~\mbox{etc},
\label{B3}
\end{eqnarray}
where $z=x-y$, it is possible to obtain from the above formulae
 \begin{eqnarray}
{\rm Tr}\hat A=A(0)\int d^2x=\int\frac{d^2p}{(2\pi)^2}
\overline{A}(p)\int d^2x.
\label{B4}
\end{eqnarray}
If there is an operator function $F(\hat A)$, where $\hat A$ is a
translationally invariant operator, then in the coordinate
representation its matrix elements depend on the difference $(x-y)$.
Obviously, it is possible to define the Fourier transformations
$\overline{F(A)}(p)$ of its matrix elements, and the following
relations are valid ($\overline{A}(p)$ is the Fourier transformation
for the matrix element $A(x-y)$):
\begin{eqnarray}
\overline{F(A)}(p)=F(\overline{A}(p));~~~
{\rm Tr}F(\hat A)=\int\frac{d^2p}{(2\pi)^2}
F(\overline{A}(p))\int d^2x.
\label{B5}
\end{eqnarray}
Finally, suppose that $\hat A$ is an operator in some internal
$n$-dimensional vector space, in addition. Evidently, the same is
valid for the Fourier transformation $\overline{A}(p)$ which is now
some $n\times n$ matrix. Let $\lambda_i(p)$ are eigenvalues of the
$n\times n$ matrix $\overline{A}(p)$, where  $i=1,2,..,n$. Then
\begin{eqnarray}
{\rm Tr}F(\hat A)=\int\frac{d^2p}{(2\pi)^2}
\mbox{tr}F(\overline{A}(p))\int
d^2x=\sum_{i=1}^{n}\int\frac{d^2p}{(2\pi)^2}
F(\lambda_i(p))\int d^2x.
\label{B6}
\end{eqnarray}
In this formula we use the notation tr for the trace of any operator
in the internal $n$-dimensional vector space only, whereas the
symbol Tr means the trace of an operator both in the coordinate and
internal spaces.

\section{Quark propagator}
\label{ApE}

It is clear from (\ref{A8}) that the quark propagator $S_0$ is the
following 2$\times$2 matrix in the two-dimensional flavor space:
\begin{eqnarray}
S_0\equiv\left (\begin{array}{cc}
S_{11}~,&S_{12} \\
 S_{21}~,&S_{22}
\end{array}\right )=\left (\begin{array}{cc}
D_{+}, & D_{12}\\
D_{21}, &D_{-}\end{array}\right )^{-1},  \label{E1}
\end{eqnarray}
where (the summation over $\alpha =0,1$ is implied)
\begin{eqnarray}
D_{\pm}=i\gamma^\alpha\partial_\alpha- M+(\mu\pm\nu)\gamma^0,~~~
D_{12}=D_{21}=-i\gamma^5\Delta. \label{E01}
\end{eqnarray}
The connection between $S_{ij}$ and $D_{ij}$ is the following:
\begin{eqnarray}
S_{11}=\left [D_{+}-D_{12}D_{-}^{-1}D_{21}\right ]^{-1},&&
S_{21}=-D_{-}^{-1}D_{21}S_{11},\nonumber\\
S_{22}=\left [D_{-}-D_{21}D_{+}^{-1}D_{12}\right ]^{-1},&&
S_{12}=-D_{+}^{-1}D_{12}S_{22}. \label{E2}
\end{eqnarray}
It is easy to establish the following relations:
\begin{eqnarray}
\label{E3} &&D_{\pm}=\int\!\frac{d^2p}{(2\pi)^2}\,
e^{-ip(x-y)}\left\{(p_0+\mu-E^\mp)\gamma^0\Lambda_++
(p_0+\mu+E^\pm)\gamma^0\Lambda_-\right\},\nonumber\\
&&(D_{\pm})^{-1}=\int\!\frac{d^2p}{(2\pi)^2}e^{-ip(x-y)}
\left\{\frac{\Lambda_+\gamma^0}{p_0+\mu-E^\mp}+
\frac{\Lambda_-\gamma^0}{p_0+\mu+E^\pm} \right\},
\end{eqnarray}
where $E^\pm=E\pm\nu$, $E=\sqrt{p_1^{~\!2}+ M^2}$,
$\Lambda_\pm=\frac 12(1\pm\frac{\gamma^0(\gamma_1 p_1+M)}{E})$, and
$\gamma_1=-\gamma^1$.
 \footnote{In (1+1)-dimensions the gamma
matrices have the form:
$$\gamma^0=\left (\begin{array}{cc}
0~,&1 \\
 1~,&0\end{array}\right ),~~\gamma^1=\left (\begin{array}{cc}
0~,&-1 \\
 1~,&0\end{array}\right ),~~\gamma^5=\left (\begin{array}{cc}
1~,&0 \\
 0~,&-1\end{array}\right ).$$}
Using the relations (\ref{E2}), it is possible to obtain from
(\ref{E01}) and (\ref{E3}):
\begin{eqnarray}
&&S_{11}=\int\!\frac{d^2p}{(2\pi)^2}e^{-ip(x-y)}
\left\{\frac{p_0+\mu+E^-}{(p_0+\mu)^2-
(E_\Delta^-)^2}\gamma^0\bar\Lambda_-+
\frac{p_0+\mu-E^+}{(p_0+\mu)^2-
(E_\Delta^+)^2}\gamma^0\bar\Lambda_+\right\},\nonumber\\
&&S_{22}=\int\!\frac{d^2p}{(2\pi)^2}e^{-ip(x-y)}
\left\{\frac{p_0+\mu+E^+}{(p_0+\mu)^2-
(E_\Delta^+)^2}\gamma^0\bar\Lambda_-+
\frac{p_0+\mu-E^-}{(p_0+\mu)^2-
(E_\Delta^-)^2}\gamma^0\bar\Lambda_+\right\},\nonumber\\
&&S_{12}=-i\Delta\int\!\frac{d^2p}{(2\pi)^2}e^{-ip(x-y)}
\left\{\frac{\gamma^5\bar\Lambda_+}{(p_0+\mu)^2- (E_\Delta^-)^2}+
\frac{\gamma^5\bar\Lambda_-}{(p_0+\mu)^2-
(E_\Delta^+)^2}\right\},\nonumber\\
&&S_{21}=-i\Delta\int\!\frac{d^2p}{(2\pi)^2}e^{-ip(x-y)}
\left\{\frac{\gamma^5\bar\Lambda_+}{(p_0+\mu)^2- (E_\Delta^+)^2}+
\frac{\gamma^5\bar\Lambda_-}{(p_0+\mu)^2- (E_\Delta^-)^2}\right\},
\label{E4}
\end{eqnarray}
where $\bar\Lambda_\pm=\frac 12(1\pm\frac{\gamma^0(\gamma_1
p_1-M)}{E})$ and $p_0$ in the integrand is a shorthand notation for
$p_0+i\varepsilon\cdot {\rm sign}(p_0)$, where $\varepsilon\to 0_+$.
This prescription for the quantity $p_0$ correctly implements the
role of the quantities $\mu$ and $\mu_I$ as the chemical potentials
and preserves the causality of the theory \cite{chod}. It is worth
also to note the following useful relations:
\begin{eqnarray*}
\gamma^5\bar\Lambda_\pm\gamma^5=\Lambda_\pm,~~~
\gamma^0\bar\Lambda_\pm\gamma^0=\Lambda_\mp.
 \label{E5}
\end{eqnarray*}
The poles of the matrix elements (\ref{E4}) of the quark propagator
in the energy-momentum space give the dispersion lows for
quasiparticles, i.e. the momentum dependence of the quark ($p_{0u}$,
$p_{0d}$) and antiquark ($p_{0\bar u}$, $p_{0\bar d}$) energies, in
a medium
\begin{equation}
p_{0u}=E_\Delta^--\mu,~~~p_{0d}=E_\Delta^+-\mu,~~
p_{0\bar u}=-(E_\Delta^++\mu),~~ p_{0\bar d}=-(E_\Delta^-+\mu).
\label{E6}
\end{equation}
Strictly speaking, the quantities $p_{0u}$, $p_{0d}$ from
(\ref{E6}) are the energies necessary for the creation of quarks
with momentum $p_1$, whereas $p_{0\bar u}$, $p_{0\bar d}$ is the
energy necessary for the annihilation of antiquarks.


\begin{thebibliography}{999}

\bibitem{njl}
Y. Nambu and G. Jona-Lasinio, Phys. Rev. D {\bf 112}, 345 (1961).

\bibitem{2}
D. Ebert and M. K. Volkov, Yad. Fiz. {\bf 36}, 1265 (1982); D. Ebert
and H. Reinhardt, Nucl. Phys. B {\bf 271}, 188 (1986).

\bibitem{asakawa}
M.~Asakawa and K.~Yazaki, Nucl.\ Phys.\ A {\bf 504}, 668 (1989);
P.~Zhuang, J.~H\"ufner and S.~P.~Klevansky, Nucl.\ Phys.\ A {\bf
576}, 525 (1994); D. Ebert, H. Reinhardt and M. K. Volkov, Prog.
Part. Nucl. Phys. {\bf 33}, 1 (1994).

\bibitem{alford}
M. Buballa, Phys. Rep. {\bf 407}, 205 (2005); I. A. Shovkovy, Found.
Phys. {\bf 35}, 1309 (2005);
 M. G.~Alford, A.~Schmitt, K.~Rajagopal, and T.~Sch\"afer,
 Rev.\ Mod.\ Phys.\  {\bf 80}, 1455 (2008).

\bibitem{klim}
  D.~Ebert, V.~V.~Khudyakov, V.~C.~Zhukovsky and K.~G.~Klimenko,
  Phys.\ Rev.\  D {\bf 65}, 054024 (2002);
D.~Ebert, K.~G.~Klimenko and V.~L.~Yudichev,
  Phys.\ Rev.\  D {\bf 75}, 045005 (2007);
  Eur.\ Phys.\ J.\  C {\bf 53}, 65 (2008);
T.~Brauner,
  Phys.\ Rev.\  D {\bf 77}, 096006 (2008);
T.~Fujihara, D.~Kimura, T.~Inagaki and A.~Kvinikhidze,
  Phys.\ Rev.\  D {\bf 79}, 096008 (2009);
G.~Lugones, A.~G.~Grunfeld, N.~N.~Scoccola and C.~Villavicencio,
  arXiv:0907.0035.

 \bibitem{warringa}
H.~J.~Warringa, D.~Boer and J.~O.~Andersen,
  Phys.\ Rev.\  D {\bf 72}, 014015 (2005);
K.~Fukushima and H.~J.~Warringa,
  Phys.\ Rev.\ Lett.\  {\bf 100}, 032007 (2008).

\bibitem{incera}
E.~J.~Ferrer, V.~de la Incera and C.~Manuel,
  Nucl.\ Phys.\  B {\bf 747}, 88 (2006);
  E.~J.~Ferrer and V.~de la Incera,
  Phys.\ Rev.\  D {\bf 76}, 045011 (2007).

\bibitem{ebert}
D.~Ebert, K.~G.~Klimenko, M.~A.~Vdovichenko and A.~S.~Vshivtsev,
  Phys.\ Rev.\  D {\bf 61}, 025005 (2000);
D.~Ebert and K.~G.~Klimenko, Nucl.\ Phys.\  A {\bf 728}, 203 (2003);
B.~Hiller, A.~A.~Osipov, A.~H.~Blin and J.~da Providencia,
  Phys.\ Lett.\  B {\bf 650}, 262 (2007);
  SIGMA {\bf 4}, 024 (2008);
E.~S.~Fraga and A.~J.~Mizher,
  Phys.\ Rev.\  D {\bf 78}, 025016 (2008);
  arXiv:0810.5162;
D.~P.~Menezes, M.~B.~Pinto, S.~S.~Avancini, A.~P.~Martinez and
  C.~Providencia,
  Phys.\ Rev.\  C {\bf 79}, 035807 (2009);
 arXiv:0907.2607;
 A.~Ayala, A.~Bashir, A.~Raya and A.~Sanchez,
  arXiv:0904.4533;
  N.~Sadooghi,
  arXiv:0905.2097.

\bibitem{vshivtsev}
A.~S.~Vshivtsev, M.~A.~Vdovichenko and K.~G.~Klimenko, J.\ Exp.\
Theor.\ Phys.\  {\bf 87}, 229 (1998);
D.~Ebert, K.~G.~Klimenko, A.~V.~Tyukov and V.~C.~Zhukovsky,
  Eur.\ Phys.\ J.\  C {\bf 58}, 57 (2008);
L.~M.~Abreu, A.~P.~C.~Malbouisson, J.~M.~C.~Malbouisson and
A.~E.~Santana,
  Nucl.\ Phys.\  B {\bf 818}, 127 (2009).

\bibitem{son}
D. T.~Son and M. A.~Stephanov, Phys.\ Atom.\ Nucl.\  {\bf 64}, 834
(2001); J. B. Kogut and D. Toublan, Phys. Rev. D {\bf 64}, 034007
(2001); J. B. Kogut and D. K. Sinclair, Phys. Rev. D {\bf 66},
014508 (2002).

\bibitem{frank}
M. Frank, M. Buballa, and M. Oertel, Phys. Lett. B {\bf 562}, 221
(2003); A. Barducci, R. Casalbuoni, G. Pettini, and L. Ravagli,
Phys. Rev. D {\bf 69}, 096004 (2004); L. He and P. Zhuang, Phys.
Lett. B {\bf 615}, 93 (2005); L. He, M. Jin, and P. Zhuang, Phys.
Rev. D {\bf 71}, 116001 (2005); T.~Herpay and P.~Kovacs,
 Phys.\ Rev.\  D {\bf 78}, 116008 (2008);
A.~A.~Andrianov and D.~Espriu,
  Phys.\ Lett.\  B {\bf 663}, 450 (2008);
A.~A.~Andrianov, V.~A.~Andrianov and D.~Espriu,
  Phys.\ Lett.\  B {\bf 678}, 416 (2009);
L.~F.~Palhares, E.~S.~Fraga and C.~Villavicencio,
  Nucl.\ Phys.\  A {\bf 820}, 287C (2009);
E.~S.~Fraga, L.~F.~Palhares and C.~Villavicencio,
  Phys.\ Rev.\  D {\bf 79}, 014021 (2009).

\bibitem{ek}
D. Ebert and K. G. Klimenko, J.\ Phys.\ G {\bf 32}, 599 (2006);
Eur.\ Phys.\ J.\  C {\bf 46}, 771 (2006).

\bibitem{jin}
L. He, M. Jin, and P. Zhuang, Phys.\ Rev.\  D {\bf 74}, 036005
(2006).

\bibitem{andersen}
J. O.~Andersen and L.~Kyllingstad, arXiv:hep-ph/0701033;
H.~Abuki, R.~Anglani, R.~Gatto, M.~Pellicoro and M.~Ruggieri,
  arXiv:0809.2658;
H.~Abuki, T.~Brauner and H.~J.~Warringa,
  arXiv:0901.2477.

\bibitem{abuki}
S.~Mukherjee, M. G.~Mustafa, and R.~Ray,
  Phys.\ Rev.\  D {\bf 75}, 094015 (2007);
H.~Abuki, M.~Ciminale, R.~Gatto, N. D.~Ippolito, G.~Nardulli, and
  M.~Ruggieri,
  Phys.\ Rev.\  D {\bf 78}, 014002 (2008);
H.~Abuki, R.~Anglani, R.~Gatto, G.~Nardulli and M.~Ruggieri,
  Phys.\ Rev.\  D {\bf 78}, 034034 (2008);
J.~O.~Andersen and T.~Brauner,
  Phys.\ Rev.\  D {\bf 78}, 014030 (2008).

\bibitem{gn}
D. J. Gross and A. Neveu, Phys. Rev. D {\bf 10}, 3235 (1974).

\bibitem{ft}
J.~Feinberg, Annals Phys.\  {\bf 309}, 166 (2004);
M.~Thies, J.\ Phys.\ A  {\bf 39}, 12707 (2006).

\bibitem{wolff}
U. Wolff, Phys. Lett. B {\bf 157}, 303 (1985); T. Inagaki, T. Kouno,
and T. Muta, Int. J. Mod. Phys. A {\bf 10}, 2241 (1995); S. Kanemura
and H.-T. Sato, Mod. Phys. Lett. A {\bf 10}, 1777 (1995).

\bibitem{kgk1}
 K. G. Klimenko, Theor.\ Math.\ Phys.\  {\bf 75}, 487 (1988).

\bibitem{barducci}
A. Barducci, R. Casalbuoni, M. Modugno, and G. Pettini, Phys. Rev. D
{\bf 51}, 3042 (1995).

\bibitem{chodos}
 A.~Chodos, H.~Minakata, F.~Cooper, A.~Singh, and W.~Mao,
  Phys. Rev. D {\bf 61}, 045011 (2000);
 K.~Ohwa, Phys.\ Rev.\  D {\bf 65}, 085040 (2002).

\bibitem{thies}
  V.~Schon and M.~Thies,
  Phys.\ Rev.\  D {\bf 62}, 096002 (2000);
 A.~Brzoska and M.~Thies,
  Phys.\ Rev.\  D {\bf 65}, 125001 (2002).

\bibitem{caldas}
A. Chodos and H. Minakata, Phys. Lett. A {\bf 191}, 39 (1994);
 H.~Caldas, J.~L.~Kneur, M.~B.~Pinto and R.~O.~Ramos,
  Phys.\ Rev.\  B {\bf 77}, 205109 (2008);
H.~Caldas, Nucl.\ Phys.\  B {\bf 807}, 651 (2009).

\bibitem{okopinska}
A. Okopinska, Phys.\ Rev.\  D {\bf 38}, 2507 (1988); S. K.~Gandhi,
H. F.~Jones and M. B.~Pinto,
  Nucl.\ Phys.\  B {\bf 359}, 429 (1991);
K. G. Klimenko, Z.\ Phys.\  C {\bf 60}, 677 (1993).
J. L.~Kneur, M. B.~Pinto, and R. O.~Ramos,
  Phys.\ Rev.\  D {\bf 74}, 125020 (2006);
  Int.\ J.\ Mod.\ Phys.\  E {\bf 16}, 2798 (2007);
A. A.~Osipov, B.~Hiller, and A. H.~Blin,
  Phys.\ Lett.\  B {\bf 653}, 346 (2007).

\bibitem{coleman}
N. D. Mermin and H. Wagner, Phys.\ Rev.\ Lett. {\bf 17}, 1133
(1966); S. Coleman, Commun. Math. Phys. {\bf 31}, 259 (1973).

\bibitem{ekzt}
D.~Ebert, K.~G.~Klimenko, A.~V.~Tyukov and V.~C.~Zhukovsky,
  Phys.\ Rev.\  D {\bf 78}, 045008 (2008).

\bibitem{kim}
S. K. Kim, W. Namgung, K. S. Soh, and J. H. Yee, Phys. Rev. D {\bf
36}, 3172 (1987); D. Y. Song and J. K. Kim, Phys. Rev. D {\bf 41},
3165 (1990); A. S. Vshivtsev, K. G. Klimenko, B. V. Magnitsky, JETP
Lett. {\bf 61}, 871 (1995);
Phys. Atom. Nucl. {\bf 59}, 529 (1996);
A. S.~Vshivtsev, A. G.~Kisun'ko, K. G.~Klimenko, and D.
V.~Peregudov, Izv. Vuz. Fiz. {\bf 41N2}, 29 (1998).

\bibitem{dunne}
G.~Basar and G.~V.~Dunne,
  Phys.\ Rev.\ Lett.\  {\bf 100}, 200404 (2008);
  Phys.\ Rev.\  D {\bf 78}, 065022 (2008);
G.~Basar, G.~V.~Dunne and M.~Thies,
  Phys.\ Rev.\  D {\bf 79}, 105012 (2009);
F.~Correa, G.~V.~Dunne and M.~S.~Plyushchay,
  arXiv:0904.2768.

\bibitem{massive}
S. Aoki and K. Higashijima, Progr. Theor. Phys. {\bf 76}, 521
(1986); J.~Feinberg and A.~Zee,
  Phys.\ Lett.\  B {\bf 411}, 134 (1997).

\bibitem{eky}
  D.~Ebert, K.~G.~Klimenko and V.~L.~Yudichev,
  Phys.\ Rev.\  C {\bf 72}, 015201 (2005);
Phys.\ Rev. D {\bf 72}, 056007 (2005);
Phys.\ Rev. D {\bf 75}, 025024 (2007).

\bibitem{chod}
A. Chodos, K. Everding and D. A. Owen, Phys. Rev. D {\bf 42}, 2881
(1990).

\bibitem{ek2}
D. Ebert and K. G. Klimenko, arXiv:0902.1861.

\bibitem{hooft}
G. `t Hooft, Nucl.Phys. B {\bf 72}, 461 (1974); D. Ebert and V. N.
Pervushin, Teor. Mat. Fiz. {\bf 36}, 313 (1978).

\bibitem{EFR}
D. Ebert, T. Feldmann and H. Reinhardt, Phys. Lett. B {\bf 388}, 154
(1996).

\end{thebibliography}
\end{document}